\documentclass[a4paper,11pt]{article}

\usepackage{latexsym}
\usepackage{amsmath}
\usepackage{a4wide}
\usepackage{graphicx, subfigure}
\usepackage{cite}
\usepackage{tabularx}
\usepackage{array}
\usepackage{graphics}
\usepackage{graphicx}
\usepackage{epsfig}                 
\usepackage{amsmath}
\usepackage{amssymb}
\usepackage{psfrag}
\usepackage{longtable}
\usepackage{verbatim}
\usepackage{cite}
\input{definitions.tex}

\begin{document}

\hfill {\tt CERN-PH-TH/2013-256,   MITP/13-064}
\begin{center}

\vspace{3.cm}

{\huge\bf On the LHCb anomaly in $B\rightarrow K^*\ell^+\ell^-$}
\vspace{2.cm}

{\Large\bf  Tobias Hurth\footnote{Email: tobias.hurth@cern.ch}
}

\vspace{0.1cm}
{\large\it PRISMA Cluster of Excellence and  Institute for Physics (THEP)\\
Johannes Gutenberg University, D-55099 Mainz, Germany}
 
\vspace{1.cm}

{\Large\bf   Farvah Mahmoudi\footnote{Email: nazila@cern.ch}                            
}

\vspace{0.1cm}
{\large\it Clermont Universit{\'e}, Universit\'e Blaise Pascal, CNRS/IN2P3,\\
LPC, BP 10448, 63000 Clermont-Ferrand, France}\\[0.2cm]
{\large\it CERN Theory Division, Physics Department,\\ CH-1211 Geneva 23, Switzerland}\\

\end{center}

\vspace{2.cm}
\thispagestyle{empty}
{\bf Abstract:} 
The latest LHCb angular analysis of the rare decay $B  \rightarrow K^{*} \mu^+\mu^-$  shows some discrepancies 
from the SM  up to the  
3.7$\sigma$ level.  There is a consistent new physics explanation of these anomalies, 
while it is also reasonable  that these anomalies are just statistical fluctuations and/or a direct consequence of 
underestimated hadronic uncertainties. We briefly discuss possible cross-checks of the various hypotheses with an analysis of the inclusive 
$B \rightarrow X_s \ell^+\ell^-$  based on the data collected by the $B$ factories Babar and Belle and also based on future  opportunities at SuperBelle.
  We also present a global analysis of the latest 
LHCb data under the hypothesis of Minimal Flavour Violation. The latter is an important benchmark 
scenario for new physics models. Any measurements beyond the MFV  bounds and relations unambiguously indicate the existence of new flavour structures next to the Yukawa couplings of the Standard Model. However, if new physics is responsible for these discrepancies, we show it is compatible with the MFV hypothesis, so no new flavour structures are needed. Moreover, we analyse the impact of the correlations between the observables based on a Monte Carlo study.

\newpage

\section{Introduction}

The recent measurements by the high-statistics  LHCb experiment \cite{LHCb} have been fully consistent with the simple Cabibbo-Kobayashi-Maskawa (CKM) 
theory of the Standard Model (SM)~\cite{Hurth:2012vp, Hurth:2010tk}.  The LHCb collaboration has not found any sizeable discrepancy in the $B$ meson sector until recently besides 
the isospin asymmetry in the rare decay mode $B \rightarrow K \mu^+\mu^ -$.
This confirms the general result  of the $B$ factories at SLAC ({BaBar} experiment) \cite{Babar}
and at KEK (Belle experiment) \cite{Belle} and of the Tevatron $B$ physics experiments
\cite{TevatronB1,TevatronB2}. 

However, more recently, with the first measurement of new angular observables in the exclusive decay $B \to K^*\mu^+\mu^-$, LHCb has announced a 3.7$\sigma$ local discrepancy 
in one of the $q^2$ bins for one of the angular observables~\cite{Aaij:2013qta}. LHCb compared here with the theoretical predictions in Ref.~\cite{Descotes-Genon:2013vna}.
In fact, as was correctly stated in Ref.~\cite{Descotes-Genon:2013wba}, the deviation is $4\sigma$ if one compares the central values of the experimental measurement and 
the theory prediction together with the corresponding error bars as it is usually done. In Refs.~\cite{Descotes-Genon:2013wba,Altmannshofer:2013foa,Gauld:2013qba,Buras:2013qja,Gauld:2013qja,Datta:2013kja,Buras:2013dea,Beaujean:2013soa} consistent SM and new physics interpretations of this deviation have been discussed.
Intriguingly, other smaller but consistent deviations are also present in other observables~\cite{Aaij:2013qta}.
  
In this paper we discuss the hadronic uncertainties, possible cross-checks using the inclusive 
$B \to X_s \ell^+\ell^-$  mode, and the impact of experimental correlations. Moreover, we investigate the possibility of new physics under the MFV hypothesis. 

This is organised as follows. In the next section we discuss the various sources of hadronic uncertainties in the exclusive mode $B \rightarrow K^* \mu^+\mu^-$ and explore the role of power corrections.
In Section 3 we work out the correlations between the measurements of the various observables. In Section 4 we present MFV-analysis of the latest data, and in Section 5 we explore 
possible cross-checks with the inclusive mode $B \to X_s \ell^+\ell^-$. The conclusions are given in Section 6.

\section{Hadronic uncertainties  in the mode $B \rightarrow K^* \mu^+\mu^-$}

\subsection{Form factor independent observables} Let us recall the construction of so-called {\it theoretically clean} angular observables in the exclusive semi-leptonic penguin mode.
The mode $B \rightarrow K^* \mu^+\mu^-$ offers a large variety of experimentally accessible observables, but the hadronic uncertainties in the theoretical predictions are in general 
large. The decay with $K^*$ on the mass shell  has a 4-fold differential distribution
\begin{equation}
  \label{eq:differential decay rate}
  \frac{d^4\Gamma[B \to K^{*}(\to K \pi)\mu\mu]}
       {d q^2\, d\ctl\, d\ctk\, d\phi} =
  \frac{9}{32\pi} \sum_i J_i(q^2)\, g_i(\theta_l, \theta_K, \phi)\,,
\end{equation}
w.r.t. the dilepton invariant mass $q^2$ and the angles $\theta_l$, $\theta_K$, 
and $\phi$ (as defined in~\cite{Egede:2008uy}).
It offers 12 observables $J_i(q^2)$, from which all other known ones 
can be derived upon integration over appropriate combinations of angles. 
The $J_i$ depend on products of the eight theoretical complex $K^*$ spin amplitudes 
$A_i$, $A_{\bot,\|,0}^{L,R},A_t,A_S$. The $J_i$ are bi-linear 
functions of the spin amplitudes such as
\begin{equation}
  \label{eq:J1s}
  J_{s}^1 = \frac{3}{4} \left[|\apeL|^2 + |\apaL|^2 + |\apeR|^2 + |\apaR|^2  \right],
\end{equation}
with the expression for the eleven other $J_i$ terms given for example 
in~\cite{Kruger:1999xa,Kruger:2005ep,Altmannshofer:2008dz,Egede:2010zc}.

In the low-$q^2$ region, the up-to-date description of exclusive heavy-to-light $B \to K^* \mu^+\mu^-$ 
decays  is the method of QCD-improved Factorisation (QCDF) and 
its field-theoretical formulation of Soft-Collinear Effective Theory 
(SCET). In the combined limit of a heavy $b$-quark and of an energetic $K^*$ meson,  the decay amplitude factorises to leading order in $\Lambda/m_b$ and to all orders in $\alpha_s$
into process-independent non-perturbative quantities like $B\to K^*$ form factors  
and light-cone distribution amplitudes (LCDAs) of the heavy (light) 
mesons and perturbatively calculable quantities, which are known to
$O(\alpha_s^1)$~\cite{Beneke:2001at,Beneke:2004dp}. 
Further, the {\it seven} a priori independent $B\to K^*$ QCD form factors reduce 
to {\it two}  universal {\it soft} form factors $\xi_{\bot,\|}$~\cite{Charles:1998dr}. 
The factorisation formula applies well in the range
of the dilepton mass range, $1\; {\rm GeV}^2 < q^2 < 6\; {\rm GeV}^2$.

Taking into account all these  simplifications the various \Kstar spin amplitudes at leading order in $\lqcd/m_b$ and \as turn out to be linear in the soft form factors $\xi_{\bot,\|}$ and also in the short-distance Wilson coefficients. As was explicitly shown in Refs.~\cite{Egede:2008uy, Egede:2010zc}, these simplifications allow  to design a set of optimised observables, in which any soft form factor dependence (and its corresponding uncertainty) cancels out for all low dilepton mass squared~\qsq at leading order in 
\as and $\lqcd/m_b$. An optimised set of independent\footnote{The number of independent observables  $J_i$ is in general smaller than 12 due to dependencies 
between them. This set of independent $J_i$ matches the number of theoretical degrees of freedom, namely the spin amplitudes $A_i$ due to symmetries of the angular distribution under specific  transformations of the $A_i$. These symmetries and relations were explicitly identified in  Refs.~\cite{Egede:2008uy,Egede:2010zc}. For the most general case this was done in Ref.~\cite{Matias:2012xw}.
However, in practice one could completely ignore these theory considerations of symmetries and relations and would recover 
them by obvious correlations between the observables.} observables  was constructed in Refs.~\cite{Matias:2012xw,Descotes-Genon:2013vna}, in which almost all  observables are free from hadronic uncertainties which are related to the form factors.

\subsection{Power corrections} 
The soft form factors  are {\it not} the only source of hadronic uncertainties in these angular observables. 
It is well-known that within the QCDF/SCET approach, a general, quantitative method to estimate the important $\lqcd/m_b$ corrections to the heavy quark limit is missing. 
In spite of the fact that the power corrections cannot be calculated, the corresponding uncertainties should be made manifest within the theory predictions. Therefore, in Refs.~\cite{Egede:2008uy,Egede:2010zc}
the effects of the $\Lambda_{\rm QCD}/m_{b}$ corrections has been parametrised for each of the \Kstarz spin-amplitudes with some {\it unknown} linear correction. In case of CP-conserving observables this just means  
\begin{equation}
A'_{i} =  A_{i}(1 + C_{i}),
\end{equation} 
where $C_{i}$ is the relative amplitude.\footnote{In the case of CP-violating observables, a strong phase has to be included (see Ref.\cite{Egede:2010zc} for details).}  It is further assumed that these amplitudes ($C_{i}$) are not functions of $q^2$, although in practise they may actually be, and any unknown correlations are  also ignored. 
An estimate of the theoretical uncertainty arising from the unknown $\Lambda_{\rm QCD}/m_{b}$ corrections can  now be made using a randomly selected ensemble. For each member of the ensemble, values of $C_i$ are chosen in the ranges $C_{i} \in [-0.1,0.1]$ or $C_{i} \in [-0.05,0.05]$ from a random uniform distribution. This is  done for the seven amplitudes, $A_t$, $A_{0}^{L,R}$,
$A_{\parallel}^{L,R}$, $A_{\perp}^{L,R}$ (neglecting the scalar amplitude), to provide a complete
description of the decay. To estimate the contribution to the theoretical uncertainties from
$\Lambda_{\rm QCD}/m_{b}$ corrections for a particular observable, each element
in the ensemble is used to calculate the value of that observable at
a fixed value of \qsq. A one $\sigma$ error is evaluated as the
interval that contains 68\% of the values around the median.
This is done for both $C_{i} \in [-0.05,0.05]$ and $C_{i} \in [-0.1,0.1]$ to
illustrate the effects of five and ten percent corrections {\it at the amplitude level}.  By
repeating this process for different values of \qsq, bands can be
built up.  
The bands illustrate the probable range for the true value of each observable, given the current central value~\cite{Egede:2010zc}. 
Some remarks are in order: 
\begin{itemize} 
\item The choice $|C_i| < 10\%$ is based on a simple dimensional
estimate. We emphasise here that there is no strict argument available
to bound the $\Lambda_{\rm QCD}/m_\b$ corrections this way.\\ 
There are {\it soft}  arguments however:
Under the assumption that the main part of the $\lqcd/m_b$ corrections is included in 
the full form factors, the difference of the theoretical
results using the full QCD form factors on one hand and the soft form factors on 
the other hand confirms this simple dimensional estimate. In fact, the comparison 
of the approaches leads to a $7\%$ shift of the central value {\it at the level of observables}. 
Secondly, one can state that
the chiral enhancement of $\Lambda_{\rm QCD}/m_\b$ corrections in the case of
hadronic $B$ decays does not occur in the case of the semileptonic
decay mode with a {\it vector} final state.
Thus, it is not expected that they are as large as 
$20-30\%$ as in the $B \to \pi\pi$ decay.
\item  The sophisticated parametrisation of the unknown $\Lambda_{\rm QCD}/m_{b}$ corrections should not hide the 
fact that this ansatz is  put in by hand and there is no rigorous theory behind this ansatz. 
In addition, it accidentally happens that these corrections cancel out in various ratios by different amount. 
Of course this simulates an effect  which  we expect also from real corrections, however, the precise features depend on the ansatz made.
\item This parametrisation of the unknown $\Lambda_{\rm QCD}/m_{b}$ was also used in all the theory predictions of Ref.~\cite{Descotes-Genon:2013vna} to which the LHCb collaboration refers.
\item  In Ref.~\cite{Jaeger} a general parametrisation for the power corrections to the form factor terms (the factorisable piece in the QCD factorisation formula) is given. But also this ansatz is just a parametrisation of our ignorance about the power corrections only. There are two free parameters in the ansatz for each QCD form factor which have to be determined. The power corrections to the non factorisable piece are here not included yet and have to be considered separately. 
\end{itemize}

\subsection{Low-recoil region}
The low-hadronic recoil region is characterised by large values of the dilepton
invariant mass $\qsq \gsim (14 - 15) \gev^2$ above the two narrow resonances of
$\jpsi$ and $\psitwos$. It is shown that local operator product expansion
is applicable ($\qsq \sim m_b^2$) \cite{Grinstein:2004vb, Beylich:2011aq} and it allows to obtain the $B \to K^* \mu^+\mu^-$ matrix element in a systematic expansion in $\alpha_s$ and in $\Lambda/m_b$. 
Most important, the leading power corrections are shown to be suppressed 
by $(\lqcd/m_b)^2$ or $\as \lqcd/m_b$ \cite{Beylich:2011aq} and to contribute 
only at the few percent level. 
The only caveat is that heavy-to-light form factors are known
only from  extrapolations from LCSR calculations at low-\qsq at present. But this is improving with direct lattice calculations in the high-$q^2$ becoming available~\cite{Horgan:2013hoa,Horgan:2013pva}.
There are improved Isgur-Wise relations between the form factors in leading power
of $\Lambda/m_b$. Their application and the introduction of specific modified Wilson coefficients
lead to simple expressions for the $K^*$ spin amplitudes to leading order in $1/m_b$ in the low
recoil region~\cite{Bobeth:2010wg,Bobeth:2011gi,Bobeth:2012vn}.

Thus, the hadronic uncertainties are well under control in the low-recoil region. But we will see below, the sensitivity to the short-distance Wilson coefficients in which also potential NP contributions enter is relatively small. 

The theoretical tools used in the low- and high-$q^2$ are different. This allows for important cross-checks in the future and might help to disentangle potential new physics signals from power corrections.

\subsection{Theory predictions and signs for new physics beyond the SM}

The LHCb collaboration reports one significant local deviation, namely in the bin $q^2 \in [ 4.3,8.63 ]$ GeV$^2$ of the observable $P_5^{'}$ within the comparison with the theory
predictions in Ref.~\cite{Descotes-Genon:2013vna}. Here LHCb states a 3.7$\sigma$ deviation~\cite{Aaij:2013qta} {\bf (i)}. 
 
All other data appear to be compatible with the SM predictions.
But comparing theory and experiment more closely, there are some other slight  deviations beyond the $2 \sigma$ level visible; in the second low-$q^2$ bin, $q^2 \in [2,4.3]$ GeV$^2$,  of $P_2$ {\bf (ii)} and in the high-$q^2$ bin, $q^2 \in [14.18,16]$ GeV$^2$, of $P_4^{'}$  {\bf (iii)}. 
 
This is also shown in a model-independent analysis given in Ref.~\cite{Descotes-Genon:2013wba}. Here NP contributions of the operators ${\cal O}_7$, ${\cal O}_9$,
${\cal O}_{10}$ and their chiral counterparts are considered in the global fit of almost all available $b \to s$ data based on the standard $\chi^2$.
The pull of the three anomalous measurements has been found to be $4 \sigma$ {\bf (i)}, $2.9 \sigma$ {\bf (ii)}, and $2.1 \sigma$ {\bf (iii)} respectively ~\cite{Descotes-Genon:2013wba}. 

It has been shown that the deviation in the observable $P_5^{\prime}$ and the small deviation in the observable $P_2$, both in the low-$q^2$ area, can be 
consistently described by a smaller 
 $C_9$ Wilson coefficient, together with a less significant contribution of a non-zero $C_9^{\prime}$ (see for example Ref.~\cite{Descotes-Genon:2013zva}). 
More recently, the authors of Ref.~\cite{Horgan:2013hoa} calculated the form factors in the low-recoil region with lattice methods and then showed that the best-fit to the 
low-recoil observables hints in the same direction as the fits to the low-$q^2$ region~\cite{Horgan:2013pva}. 
This consistency is quite remarkable, since different theory methods are used in the two kinematical regions.

However, there are also some critical remarks in order, specifically on the largest deviation related to the observable $P_5^{'}$:

\begin{itemize}
\item The uncertainties due to power corrections in Ref.~\cite{ Descotes-Genon:2013vna} should just make these unknown corrections manifest 
and are therefore separately given in the tables of that publication.  
The procedure given above leads often accidentally to very small uncertainties of 3-5$\%$ at the observable level. 
This might be an underestimation of the hadronic uncertainty. However, if we   assume $10\%$ error due to the unknown power corrections - 
which corresponds to a naive dimension estimate of $\Lambda/m_b$ and is also backed up by some soft arguments  (see above) - we find the pull in case of the third bin of the observable $P_5^{'}$ reduced from $4.0\sigma$ to $3.6\sigma$ what still represents a significant deviation.  And even if one assumes $30\%$ error then the pull in this case is still $2.2\sigma$ within the model-independent analysis presented in Ref.~\cite{Descotes-Genon:2013wba}.

\item The validity of the QCD factorisation approach within the region $q^2 \in [ 4.3,8.63 ]$ GeV$^2$ is highly questionable. 
The validity is commonly assumed up to 6 GeV$^2$ where the kinematical assumptions about the large energy of the $K^*$ is still reasonable. Thus, using the theory of SCET/QCD factorisation up to 8.63 GeV$^2$ could induce larger hadronic corrections.

\item  Only using the measurement of the integrated $q^2 \in [1,6]$ GeV$^2$ as done in Ref.\cite{Altmannshofer:2013foa,Beaujean:2013soa}
circumvents this problem. The LHCb collaboration has presented also this measurement and states a $2.5\sigma$ deviation with respect to the SM.\cite{Aaij:2013qta} 
The integration over the complete low-$q^2$ region also smears out the potential 
new physics signals. But it is the $q^2$-dependence which might be crucial for the new physics signal to be visible.
Clearly, averaging over the full low-$q^2$ bin will often lead to a smaller deviation from the SM. This could explain the reduced discrepancy in this bin found by the LHCb collaboration.

\item There is another issue, namely the role of the charm-loop effects which were tackled in Ref.\cite{Khodjamirian:2010vf}.
In Ref.~\cite{Descotes-Genon:2013wba} it is argued that these contributions tend to enhance the new physics signal due to their specific sign. But in Ref.~\cite{Khodjamirian:2010vf} only soft-gluon contributions were considered via an OPE which is valid below the charm threshold only. Thus, a model-dependent extrapolation to higher $q^2$ via a dispersion relation  is needed. And spectator contributions were not considered yet, so the sign of the {\it complete} non-perturbative charm effects is not fixed yet and could change.

\item  We should also mention the contributions of the $K\pi$ system in an S-wave configuration. The presence of such background would pollute the angular distributions and bias the measurement of the observables~\cite{Becirevic:2012dp,Matias:2012qz}. The size of the S-wave component in the $K^*$ mass window is difficult to estimate from the theoretical point of view. Possible implications of neglecting this contribution have been discussed in Ref.~\cite{Blake:2012mb}. At present, these effects are difficult to predict, but they are taken into account in the experimental analysis and added as systematics. LHCb has set an upper limit on the contribution of the S-wave in their $K\pi$ mass window~\cite{Aaij:2013qta} by exploiting the interference between P- and S-wave and using the change of phase of the P-wave in the pole of the Breit-Wigner. Systematics due to the interference terms have been taken into account by using the bounds derived in Ref.~\cite{Matias:2012qz}. These systematics are rather small compared to the statistical error. An explanation of the anomaly in terms of interference with an S-wave system seems at the moment unlikely.

\end{itemize}

\section{Experimental fit correlations}
\label{sec:exp}

The LHCb experiment uses particular folding techniques to access the observables of interest. 
This procedure largely breaks experimental correlations between the different observables. 
In order to investigate the residual correlations, a \textit{toy} Monte Carlo study with simulated pseudo-experiment was performed~\cite{Serra}. 
Several datasets with the same number of signal events observed
by LHCb in each bin of $q^2$ are generated with the full angular probability density function of $B^0\to K^*\mu^+\mu^-$. The observables are generated around the measured values by LHCb in Refs.~\cite{Aaij:2013iag,Aaij:2013qta}, and the described folding techniques are applied to each datasets.
The eight angular observables are then extracted  with an unbinned likelihood fit, obtaining eight values for each dataset. 
The correlation coefficient is then computed assuming linear correlations among the different observables.
The correlation matrix is shown for the $q^2$ bin $[4.3,8.68]$ GeV$^2$ in Table~\ref{tab:correlations_bin3}. The other correlation matrices can be found in the Appendix. It is important to note that this correlation matrix does not contain information about the correlation due to the background or due to systematic uncertainties, which cannot be evaluated with a toy Monte Carlo study. 
The main motivation of this study is to investigate the correlation of the fitting procedure after folding. 
\begin{table}[!h]
\begin{center}
\begin{tabular}{|c|cccccccc|}
\hline
& $P_1$ & $P_2$ & $P_4^{\prime}$ & $P_5^{\prime}$ & $P_6^{\prime}$  & $P_8^{\prime}$ &$F_L$ & $A_{FB}$\\
\hline
$P_1$ & 1.00 & -0.02 & 0.14 & -0.03 & -0.03 & 0.04 & 0.07 & 0.02 \\ 
$P_2$ & -0.02 & 1.00 & 0.03 & 0.18 & -0.07 & -0.02 & -0.13 & -0.97 \\ 
$P_4^{\prime}$ & 0.14 & 0.03 & 1.00 & -0.16 & -0.05 & 0.03 & -0.04 & -0.03 \\ 
$P_5^{\prime}$ & -0.03 & 0.18 & -0.16 & 1.00 & 0.04 & 0.01 & 0.02 & -0.18 \\ 
$P_6^{\prime}$ & -0.03 & -0.07 & -0.05 & 0.04 & 1.00 & -0.14 & -0.01 & 0.07 \\ 
$P_8^{\prime}$ & 0.04 & -0.02 & 0.03 & 0.01 & -0.14 & 1.00 & 0.01 & 0.02 \\ 
$F_L$ & 0.07 & -0.13 & -0.04 & 0.02 & -0.01 & 0.01 & 1.00 & 0.13 \\ 
$A_{FB}$ & 0.02 & -0.97 & -0.03 & -0.18 & 0.07 & 0.02 & 0.13 & 1.00 \\ 
\hline
\end{tabular}
\caption{Correlation matrix for the $q^2$ region $[4.3,8.68]$ GeV$^2$ estimated by
using a toy Monte Carlo technique~\cite{Serra}.\label{tab:correlations_bin3}} 
\end{center}
\end{table}

The correlation matrix includes both
$A_{FB}$ and $P_2$. In this case the same Pdf is used to fit the folded dataset, by using the relation $A_{FB}=-\frac{3}{2}(1-F_L)P_2$. 
As expected, we found that these observables have a correlation exceeding 90\% in most of the bins. For this reason we prefer to use the observable $F_{L}$, which 
does not exhibit such a strong correlation with $P_2$, in place of $A_{FB}$. 
{All experimental measurements of the other decays} used in our fit are assumed to be independent.
A covariance matrix is built using the correlation matrices and it is used to compute the $\chi^2$ probability with each NP scenario. 
Both theoretical and experimental errors are assumed to be independent in the different bins. 
It has been checked that the impact of these correlations in the MFV analysis is small, as expected since the correlation matrices, after excluding $A_{FB}$, are
almost diagonal.\footnote{It is clear that including or not the correlations would make a significant difference if $A_{FB}$ were used instead of $F_{L}$, as done in Ref.~\cite{Descotes-Genon:2013wba}.}
However, for completeness these matrices are included in the analysis presented in the next section.

\section{General MFV analysis} 

\subsection{MFV hypothesis}
It is not easy to find a concrete NP model which is consistent with the LHCb anomaly~\cite{Gauld:2013qja}.
However, assuming that the LHCb anomaly is a hint for NP, the question if new flavour structures are needed or not is an obvious one.  

The hypothesis of  MFV~\cite{Chivukula:1987py,Hall:1990ac,D'Ambrosio:2002ex,Hurth:2008jc,Hurth:2012jn}, 
implies that flavour and CP symmetries are broken as in the SM. Thus, it 
requires that all flavour- and CP-violating interactions be  linked to the known structure of Yukawa couplings.     
The MFV hypothesis represents an important benchmark in the sense that any measurement  which 
is inconsistent with the general constraints and relations induced by the MFV 
hypothesis unambiguously indicates the existence of new flavour structures. 

Moreover, compared with a general model-independent analysis,
the number of free parameters is heavily reduced due to the additional MFV relations. 
Our analysis is based on the MFV effective Hamiltonian relevant to $b \to s$ transitions:

\begin{eqnarray}
{\cal H}^{ b\to s}_{\rm eff} &=& -\frac{4 G_F}{\sqrt{2}}
\Bigl\lbrace \bigl[  V^*_{us} V_{ub} (C^c_1 P^u_1 + C^c_2 P^u_2)
  + V^*_{cs} V_{cb} (C^c_1 P^c_1 + C^c_2 P^c_2) \bigr]
\nonumber \\
&& + { \sum_{i=3}^{10} \bigl[(V^*_{us} V_{ub}
+ V^*_{cs} V_{cb}) C^c_i \; + \; V^*_{ts} V_{tb} C^t_i \bigr] P_i +
V^*_{ts} V_{tb}
C^\ell_{0} P^\ell_{0}}~+~{\rm h.c.}  \Bigr\rbrace
\label{eq:newHeff}
\end{eqnarray}
with
\begin{equation}
\begin{array}{ll}
P^u_1 =  (\bar{s}_L \gamma_{\mu} T^a u_L) (\bar{u}_L \gamma^{\mu}
T^a b_L)~, ~~~~~&
\vspace*{0.3cm}
P_5 =  (\bar{s}_L \gamma_{\mu_1}
                   \gamma_{\mu_2}
                   \gamma_{\mu_3}    b_L)\sum_q (\bar{q} \gamma^{\mu_1}
                                                         \gamma^{\mu_2}
                                                         \gamma^{\mu_3}     q)~,
 \\
P^u_2 =  (\bar{s}_L \gamma_{\mu}     u_L) (\bar{u}_L \gamma^{\mu}
b_L)~, & \vspace*{0.3cm}
P_6 =  (\bar{s}_L \gamma_{\mu_1}
                   \gamma_{\mu_2}
                   \gamma_{\mu_3} T^a b_L)\sum_q (\bar{q} \gamma^{\mu_1}
                                                          \gamma^{\mu_2}
                                                          \gamma^{\mu_3} T^a q)~,\\
P^c_1 =  (\bar{s}_L \gamma_{\mu} T^a c_L) (\bar{c}_L \gamma^{\mu}
T^a b_L)~,
& \vspace*{0.3cm}
P_7  =   \frac{e}{16\pi^2} m_b (\bar{s}_L \sigma^{\mu \nu}     b_R)
F_{\mu \nu}~,\\
P^c_2 =  (\bar{s}_L \gamma_{\mu}     c_L) (\bar{c}_L \gamma^{\mu}
b_L)~,
&\vspace*{0.3cm}
P_8  =   \frac{g_s}{16\pi^2} m_b (\bar{s}_L \sigma^{\mu \nu} T^a b_R)
G_{\mu \nu}^a~,\\
P_3 =  (\bar{s}_L \gamma_{\mu}     b_L) \sum_q (\bar{q}\gamma^{\mu}
q)~,
&\vspace*{0.3cm}
P_9  =   \frac{e^2}{16\pi^2} (\bar{s}_L \gamma_{\mu} b_L) \sum_\ell
 (\bar{\ell}\gamma^{\mu} \ell)~,\\
P_4 =  (\bar{s}_L \gamma_{\mu} T^a b_L) \sum_q (\bar{q}\gamma^{\mu}
T^a q)~,
&\vspace*{0.3cm}
P_{10} =  \frac{e^2}{16\pi^2} (\bar{s}_L \gamma_{\mu} b_L) \sum_\ell
                             (\bar{\ell} \gamma^{\mu} \gamma_5 \ell)~.\\
\end{array}
\end{equation}
In addition we have the following scalar-density operator with right-handed $b$-quark
\begin{equation}
P^\ell_{0} = \frac{e^2}{16\pi^2} (\bar s_L b_R) (\bar \ell_R \ell_L)~.
\end{equation} 
Following our previous analyses~\cite{Hurth:2008jc,Hurth:2012jn}, we leave out the four-quark operators $P_{1-6}$  
because most of the NP contributions to them could be reabsorbed into the Wilson coefficients of the FCNC operators. 
The NP contributions are parametrised as usual:
\begin{equation}
\delta C_i(\mu_b) = C_i^{\rm MFV}(\mu_b) - C_i^{\rm SM}(\mu_b)\;.
\end{equation}
where the $C_i^{\rm SM}(\mu_b)$ are given in Table~\ref{tab:wilson}.
\begin{table}
\begin{center}
\footnotesize{\begin{tabular}{|c|c|c|c|c|}\hline
$C_7^{\text{eff}}(\mu_b)$ & $C_8^{\text{eff}}(\mu_b)$ & $C_9(\mu_b)$ & $C_{10}(\mu_b)$ & $C_0^\ell(\mu_b)$  \\ \hline
-0.2974  & -0.1614  & 4.2297 & -4.2068 & 0  \\ \hline
\end{tabular}}
\caption{SM Wilson coefficients at $\mu_b=m_b^{\text{pole}}$ and $\mu_0=2M_W$ to NNLO accuracy in $\alpha_s$. \label{tab:wilson}}
\end{center}
\end{table}

\subsection{Numerical details}

Compared to the analysis in Ref.~\cite{Hurth:2012jn}
we have the following three main changes within the experimental input: 
\begin{itemize}
\item We include now the complete new dataset on $B \to K^* \mu^+ \mu^-$ from Ref.~\cite{Aaij:2013qta}.
\item We use the new average of the $B_s\to\mu^+\mu^-$ measurement of~\cite{ Aaij:2013aka,Chatrchyan:2013bka,CMSandLHCbCollaborations:2013pla}. 
\item We take into account the experimental correlations between the $B \to K^* \mu^+ \mu^-$ observables as described in Section~\ref{sec:exp}. 
\end{itemize}
We have used 
the input parameters of Table~\ref{tab:input} and the program {\tt SuperIso v3.4}~\cite{Mahmoudi:2007vz,Mahmoudi:2008tp}
in order to obtain the theoretical predictions. 
\begin{table}[!h]
\begin{center}
\footnotesize{\begin{tabular}{|lr|lr|}\hline
$m_B=5.27917$ GeV                         & \cite{Beringer:1900zz}     &        $m_{B_s} = 5.36677 $ GeV& \cite{Beringer:1900zz}                            \\
$m_{K^*}=0.89594$ GeV                     & \cite{Beringer:1900zz}     & $|V_{tb}V_{ts}^*|=0.0401 ^{+0.0011}_{-0.0007}$         & \cite{Beringer:1900zz}          \\ \hline
$m_b^{\overline{MS}}(m_b)=4.18 \pm 0.03$ GeV & \cite{Beringer:1900zz}     & $m_c^{\overline{MS}}(m_c)=1.275 \pm 0.025$ GeV   & \cite{Beringer:1900zz}\\ 
$m_t^{pole}=173.5 \pm0.6 \pm0.8$ GeV       & \cite{Beringer:1900zz}     &$m_{\mu}=0.105658$ GeV                    & \cite{Beringer:1900zz} \\ \hline  
$\alpha_s(M_Z)=0.1184 \pm 0.0007$         & \cite{Beringer:1900zz}     & $\hat \alpha_{em}(M_Z)=1/127.916 $                     & \cite{Beringer:1900zz}          \\ 
$\alpha_s(\mu_b)=0.2161$                  &                          &$\hat\alpha_{em}(m_b)=1/133$                               &           \\ 
$\sin^2\hat\theta_W(M_Z)=0.23116(13)$     & \cite{Beringer:1900zz}&$G_F/(\hbar c)^3=1.16637(1)\;\textrm{GeV}^{-2}$& \cite{Beringer:1900zz}\\ \hline
$ f_B=194 \pm 10$ MeV                          & \cite{Hurth:2012jn}& $\tau_B=1.519 \pm0.007$ ps                             & \cite{Beringer:1900zz}          \\
$ f_{B_s} = 234 \pm 10 {\rm MeV}$ & \cite{Hurth:2012jn} & $ \tau_{B_s} = 1.497 \pm 0.026\ {\rm ps}    $ & \cite{Beringer:1900zz} \\
 \hline
$f_{K^*,\perp}$(1 GeV)$=0.185 \pm0.009$ GeV  & \cite{Ball:2007zt}         & $f_{K^*,\parallel}=0.220 \pm0.005$ GeV                 & \cite{Ball:2007zt}              \\
$a_{1,\perp}$(1 GeV)$=0.10\pm0.07$          & \cite{Ball:2004rg}       & $a_{1,\parallel}$(1 GeV)$=0.10 \pm0.07$                   & \cite{Ball:2004rg}              \\
$a_{2,\perp}$(1 GeV)$=0.13 \pm0.08$          & \cite{Ball:2004rg}       & $a_{2,\parallel}$(1 GeV)$=0.09 \pm0.05$                   & \cite{Ball:2004rg}              \\
$\lambda_{B,+}$(1 GeV)$=0.46 \pm 0.11$ GeV   & \cite{Ball:2006nr}       &  &             \\ \hline
$\mu_b=m_b^{pole}$                        &                          & $\mu_0=2 M_W$                                          &                               \\ 
$\mu_f=\sqrt{0.5 \times \mu_b}$ GeV       & \cite{Beneke:2004dp}     &                                                        &                               \\ \hline
\end{tabular}}
\caption{Input parameters. \label{tab:input}}
\end{center}
\end{table}
  
{The set of observables used in this study are provided in Table~\ref{tab:obs}, together with the SM predictions and the experimental results. To investigate the allowed regions of the Wilson coefficients in view of the current measurements}, we scan over $\delta C_7$, $\delta C_8$, $\delta C_9$, $\delta C_{10}$ and $\delta C_0^\ell$ at the $\mu_b$ scale. For each point, we then compute the flavour observables and compare with the experimental results by calculating $\chi^2$ as:
\begin{eqnarray}
\chi^2\displaystyle &=&
\sum_{\rm bins} \quad \Bigl[\sum_{j,k \in ({B\to K^* \mu^+ \mu^- \,{\rm obs.}})} (O_j^{\rm exp} - O_j^{\rm th}) \, (\sigma^{({\rm bin})})^{-1}_{jk} \,  (O_k^{\rm exp} - O_k^{\rm th}) \Bigr] \nonumber\\
&& + \sum_{i\in ({\rm other \; obs.})} \frac{(O_i^{\rm exp} - O_i^{\rm th})^2}{(\sigma_i^{\rm exp})^2 + (\sigma_i^{\rm th})^2} 
\;,\label{eq:chi2abs}
\end{eqnarray}
where $O_i^{\rm exp}$ and $O_i^{\rm th}$ are the central values of the
experimental result and theoretical prediction of observable $i$
respectively. The first term is the contribution to the $\chi^2$ from
the $B\to K^* \mu^+ \mu^-$ observables including the experimental
correlations. The $(\sigma^{({\rm bin})})^{-1}$ are the inverse of the
covariance matrices for each bin, computed using the correlations given in the Appendix. The
second term is a $\chi^2$, quantifying the agreement between
theory predictions and experimental measurements without correlations,
using all the other observables, $\sigma_i^{\rm exp}$ and
$\sigma_i^{\rm th}$ being their experimental and theoretical errors
respectively. The global fits are obtained by minimisation of the
$\chi^2$. 

We do not consider the difference of the $\chi^2$ with the minimum $\chi^2$, but
directly obtain the allowed regions from the absolute $\chi^2$ computed using Eq.~(\ref{eq:chi2abs}). This procedure leads to larger allowed
regions with respect to the use of the $\Delta \chi^2$. 
This is due to the fact that some of the observables are less
sensitive to some Wilson coefficients, while they contribute in a
democratic way to the number of degrees of freedom. 
The statistical meaning of the two dimensional contours is that for a point in the 1$\sigma$ interval allowed region, there is at least one solution with the corresponding
values of the Wilson coefficients in MFV that has a $\chi^2$ probability
corresponding to less than one Gaussian standard deviation with respect to the
full set of measurements. 
Using this method is justified since we are not aiming to determine a preferred direction to which the current results with the observed anomalies would lead, but instead our goal is to examine the global agreement of the data with the MFV predictions.

It is important to note that the exclusion plots in our MFV analysis presented in  the following section cannot be directly compared with the ones of the model-independent analyses in Refs.~\cite{Descotes-Genon:2013wba,Altmannshofer:2013foa,Beaujean:2013soa}. The main reason is that the operator basis of the MFV analysis used here is different from the set adopted in those analyses (see previous subsection). Another reason is mentioned above, namely that we use the absolute $\chi^2$ to derive the allowed regions.

\begin{table}
\begin{center}
\footnotesize{\begin{tabular}{|l|l|l|}\hline
Observable & Experiment & SM prediction \\ \hline
BR($B \to X_s \gamma$) & $(3.43 \pm 0.21\pm 0.07)\times 10^{-4}$ & $(3.09 \pm 0.24)\times 10^{-4}$\\
$\Delta_0(B \to K^* \gamma)$ & $(5.2 \pm 2.6)\times 10^{-2}$ & $(7.9 \pm 3.9)\times 10^{-2}$\\
BR($B \to X_d \gamma$) & $(1.41 \pm 0.57)\times 10^{-5}$ & $(1.49 \pm 0.30)\times 10^{-5}$ \\
BR($B_s \to \mu^+\mu^-$) & $(2.9 \pm 0.7)\times 10^{-9}$ & $(3.49 \pm 0.38) \times 10^{-9}$\\
BR($B_d \to \mu^+\mu^-$) & $(3.6 \pm 1.6)\times 10^{-10}$ & $(1.07 \pm 0.27) \times 10^{-10}$\\
BR($B \to X_s \ell^+\ell^-)_{q^2\in[1,6] \rm{GeV}^2}$ & $(1.60 \pm 0.68)\times 10^{-6}$ & $(1.73 \pm 0.16)\times 10^{-6}$\\
BR($B \to X_s \ell^+\ell^-)_{q^2>14.4 \rm{GeV}^2}$ & $(4.18 \pm 1.35)\times 10^{-7}$ &  $(2.20 \pm 0.44)\times 10^{-7}$\\
\hline
$\langle dBR/dq^2(B \to K^* \mu^+\mu^-) \rangle_{q^2\in[0.1,2] \rm{GeV}^2}$ & $(0.60 \pm 0.06 \pm 0.05 \pm 0.04 \pm 0.05 )\times 10^{-7}$ & $(0.70 \pm 0.81 )\times 10^{-7}$\\
$\langle F_{L}(B \to K^* \mu^+ \mu^-) \rangle_{q^2\in[0.1,2] \rm{GeV}^2}$ & $ 0.37 \pm 0.10 \pm 0.04$ & $ 0.32 \pm 0.20 $\\
$\langle P_1(B \to K^* \mu^+\mu^-) \rangle_{q^2\in[0.1,2] \rm{GeV}^2}$ & $ -0.19 \pm 0.40 \pm 0.02 $ & $ -0.01 \pm 0.04 $\\
$\langle P_2(B \to K^* \mu^+\mu^-) \rangle_{q^2\in[0.1,2] \rm{GeV}^2}$ & $ 0.03 \pm 0.15 \pm 0.01$ & $ 0.17 \pm 0.02 $\\
$\langle P_4'(B \to K^* \mu^+\mu^-) \rangle_{q^2\in[0.1,2] \rm{GeV}^2}$ & $0.00 \pm 0.52 \pm 0.06$ & $ -0.37 \pm 0.03 $\\
$\langle P_5'(B \to K^* \mu^+\mu^-) \rangle_{q^2\in[0.1,2] \rm{GeV}^2}$ & $0.45 \pm 0.22 \pm 0.09$ & $ 0.52 \pm 0.04 $\\
$\langle P_6'(B \to K^* \mu^+\mu^-) \rangle_{q^2\in[0.1,2] \rm{GeV}^2}$ & $0.24 \pm 0.22 \pm 0.05$ & $ -0.05 \pm 0.04 $\\
$\langle P_8'(B \to K^* \mu^+\mu^-) \rangle_{q^2\in[0.1,2] \rm{GeV}^2}$ & $-0.12 \pm 0.56 \pm 0.04$ & $ 0.02 \pm 0.04 $\\
\hline
$\langle dBR/dq^2(B \to K^* \mu^+\mu^-) \rangle_{q^2\in[2,4.3] \rm{GeV}^2}$ & $(0.30 \pm 0.03 \pm 0.03 \pm 0.02 \pm 0.02)\times 10^{-7}$ & $( 0.35 \pm 0.29 )\times 10^{-7}$\\
$\langle F_{L}(B \to K^* \mu^+ \mu^-) \rangle_{q^2\in[2,4.3] \rm{GeV}^2}$ & $ 0.74 \pm 0.10 \pm 0.03$ & $ 0.76 \pm 0.20$\\
$\langle P_1(B \to K^* \mu^+\mu^-) \rangle_{q^2\in[2,4.3] \rm{GeV}^2}$ & $ -0.29 \pm 0.65 \pm 0.03$ & $ -0.05 \pm 0.05 $\\
$\langle P_2(B \to K^* \mu^+\mu^-) \rangle_{q^2\in[2,4.3] \rm{GeV}^2}$ & $ 0.50 \pm 0.08 \pm 0.02 $ & $ 0.25 \pm 0.09 $\\
$\langle P_4'(B \to K^* \mu^+\mu^-) \rangle_{q^2\in[2,4.3] \rm{GeV}^2}$ & $0.74 \pm 0.58 \pm 0.16$ & $ 0.54 \pm 0.07 $\\
$\langle P_5'(B \to K^* \mu^+\mu^-) \rangle_{q^2\in[2,4.3] \rm{GeV}^2}$ & $0.29 \pm 0.39 \pm 0.07$ & $ -0.33 \pm 0.11 $\\
$\langle P_6'(B \to K^* \mu^+\mu^-) \rangle_{q^2\in[2,4.3] \rm{GeV}^2}$ & $-0.15 \pm 0.38 \pm 0.05$ & $ -0.06 \pm 0.06 $\\
$\langle P_8'(B \to K^* \mu^+\mu^-) \rangle_{q^2\in[2,4.3] \rm{GeV}^2}$ & $-0.3 \pm 0.58 \pm 0.14$ & $ 0.04 \pm 0.05 $\\
\hline
$\langle dBR/dq^2(B \to K^* \mu^+\mu^-) \rangle_{q^2\in[4.3,8.68] \rm{GeV}^2}$ & $(0.49 \pm 0.04 \pm 0.04 \pm 0.03 \pm 0.04)\times 10^{-7}$ & $(0.48 \pm 0.53)\times 10^{-7}$\\
$\langle F_{L}(B \to K^* \mu^+ \mu^-) \rangle_{q^2\in[4.3,8.68] \rm{GeV}^2}$ & $ 0.57 \pm 0.07 \pm 0.03$ & $ 0.63 \pm 0.14$\\
$\langle P_1(B \to K^* \mu^+\mu^-) \rangle_{q^2\in[4.3,8.68] \rm{GeV}^2}$ & $ 0.36 \pm 0.31 \pm 0.03$ & $ -0.11 \pm 0.06 $\\
$\langle P_2(B \to K^* \mu^+\mu^-) \rangle_{q^2\in[4.3,8.68] \rm{GeV}^2}$ & $ -0.25 \pm 0.08 \pm 0.02$ & $ -0.36 \pm 0.05$\\
$\langle P_4'(B \to K^* \mu^+\mu^-) \rangle_{q^2\in[4.3,8.68] \rm{GeV}^2}$ & $1.18 \pm 0.30 \pm 0.10$ & $ 0.99 \pm 0.03 $\\
$\langle P_5'(B \to K^* \mu^+\mu^-) \rangle_{q^2\in[4.3,8.68] \rm{GeV}^2}$ & $-0.19 \pm 0.16 \pm 0.03$ & $ -0.83 \pm 0.05 $\\
$\langle P_6'(B \to K^* \mu^+\mu^-) \rangle_{q^2\in[4.3,8.68] \rm{GeV}^2}$ & $0.04 \pm 0.15 \pm 0.05$ & $ -0.02 \pm 0.06 $\\
$\langle P_8'(B \to K^* \mu^+\mu^-) \rangle_{q^2\in[4.3,8.68] \rm{GeV}^2}$ & $0.58 \pm 0.38 \pm 0.06$ & $ 0.02 \pm 0.06 $\\
\hline
$\langle dBR/dq^2(B \to K^* \mu^+\mu^-) \rangle_{q^2\in[14.18,16] \rm{GeV}^2}$ & $(0.56 \pm 0.06 \pm 0.04 \pm 0.04 \pm 0.05)\times 10^{-7}$ & $(0.67 \pm 1.17)\times 10^{-7}$\\
$\langle F_{L}(B \to K^* \mu^+ \mu^-) \rangle_{q^2\in[14.18,16] \rm{GeV}^2}$ & $ 0.33 \pm 0.08 \pm 0.03$ & $ 0.39 \pm 0.24 $\\
$\langle P_1(B \to K^* \mu^+\mu^-) \rangle_{q^2\in[14.18,16] \rm{GeV}^2}$ & $ 0.07 \pm 0.28 \pm 0.02$ & $ -0.32 \pm 0.70$\\
$\langle P_2(B \to K^* \mu^+\mu^-) \rangle_{q^2\in[14.18,16] \rm{GeV}^2}$ & $ -0.50 \pm 0.03 \pm 0.01$ & $ -0.47 \pm 0.14$\\
$\langle P_4'(B \to K^* \mu^+\mu^-) \rangle_{q^2\in[14.18,16] \rm{GeV}^2}$ & $-0.18 \pm 0.70 \pm 0.08$ & $ 1.15 \pm 0.33$\\
$\langle P_5'(B \to K^* \mu^+\mu^-) \rangle_{q^2\in[14.18,16] \rm{GeV}^2}$ & $-0.79 \pm 0.20 \pm 0.18$ & $ -0.82 \pm 0.36$\\
$\langle P_6'(B \to K^* \mu^+\mu^-) \rangle_{q^2\in[14.18,16] \rm{GeV}^2}$ & $0.18 \pm 0.25 \pm 0.03$ & $ 0.00 \pm 0.00$\\
$\langle P_8'(B \to K^* \mu^+\mu^-) \rangle_{q^2\in[14.18,16] \rm{GeV}^2}$ & $-0.40 \pm 0.60 \pm 0.06$ & $ 0.00 \pm 0.01$\\
\hline
$\langle dBR/dq^2(B \to K^* \mu^+\mu^-) \rangle_{q^2\in[16,19] \rm{GeV}^2}$ & $(0.41 \pm 0.04 \pm 0.04 \pm 0.03 \pm 0.03)\times 10^{-7}$ & $(0.43 \pm 0.78)\times 10^{-7}$\\
$\langle F_{L}(B \to K^* \mu^+ \mu^-) \rangle_{q^2\in[16,19] \rm{GeV}^2}$ & $ 0.38 \pm 0.09 \pm 0.03$ & $ 0.36 \pm 0.13 $\\
$\langle P_1(B \to K^* \mu^+\mu^-) \rangle_{q^2\in[16,19] \rm{GeV}^2}$ & $ -0.71 \pm 0.35 \pm 0.06$ & $ -0.55 \pm 0.59 $\\
$\langle P_2(B \to K^* \mu^+\mu^-) \rangle_{q^2\in[16,19] \rm{GeV}^2}$ & $ -0.32 \pm 0.08 \pm 0.01$ & $ -0.41 \pm 0.15 $\\
$\langle P_4'(B \to K^* \mu^+\mu^-) \rangle_{q^2\in[16,19] \rm{GeV}^2}$ & $0.70 \pm 0.52 \pm 0.06$ & $ 1.24 \pm 0.25 $\\
$\langle P_5'(B \to K^* \mu^+\mu^-) \rangle_{q^2\in[16,19] \rm{GeV}^2}$ & $-0.60 \pm 0.19 \pm 0.09$ & $ -0.66 \pm 0.37 $\\
$\langle P_6'(B \to K^* \mu^+\mu^-) \rangle_{q^2\in[16,19] \rm{GeV}^2}$ & $-0.31 \pm 0.38 \pm 0.10$ & $ 0.00 \pm 0.00 $\\
$\langle P_8'(B \to K^* \mu^+\mu^-) \rangle_{q^2\in[16,19] \rm{GeV}^2}$ & $0.12 \pm 0.54 \pm 0.04$ & $ 0.00 \pm 0.04 $\\
\hline
$\langle dBR/dq^2(B \to K^* \mu^+\mu^-) \rangle_{q^2\in[1,6] \rm{GeV}^2}$ & $(0.34 \pm 0.03 \pm 0.04 \pm 0.02 \pm 0.03)\times 10^{-7}$ & $(0.38 \pm 0.33)\times 10^{-7}$\\
$\langle F_{L}(B \to K^* \mu^+ \mu^-) \rangle_{q^2\in[1,6] \rm{GeV}^2}$ & $ 0.65 \pm 0.08 \pm 0.03$ & $ 0.70 \pm 0.21 $\\
$\langle P_1(B \to K^* \mu^+\mu^-) \rangle_{q^2\in[1,6] \rm{GeV}^2}$ & $ 0.15 \pm 0.41 \pm 0.03$ & $ -0.06 \pm 0.04 $\\
$\langle P_2(B \to K^* \mu^+\mu^-) \rangle_{q^2\in[1,6] \rm{GeV}^2}$ & $ 0.33 \pm 0.12 \pm 0.02$ & $ 0.10 \pm 0.08 $\\
$\langle P_4'(B \to K^* \mu^+\mu^-) \rangle_{q^2\in[1,6] \rm{GeV}^2}$ & $0.58 \pm 0.36 \pm 0.06$ & $ 0.53 \pm 0.07 $\\
$\langle P_5'(B \to K^* \mu^+\mu^-) \rangle_{q^2\in[1,6] \rm{GeV}^2}$ & $0.21 \pm 0.21 \pm 0.03$ & $ -0.34 \pm 0.10 $\\
$\langle P_6'(B \to K^* \mu^+\mu^-) \rangle_{q^2\in[1,6] \rm{GeV}^2}$ & $0.18 \pm 0.21 \pm 0.03$ & $ -0.05 \pm 0.05 $\\
$\langle P_8'(B \to K^* \mu^+\mu^-) \rangle_{q^2\in[1,6] \rm{GeV}^2}$ & $0.46 \pm 0.38 \pm 0.04$ & $ 0.03 \pm 0.04$\\
\hline
\end{tabular}}
\caption{Input observables: The experimental data represent the most recent ones.  
The updated SM predictions are based on the input parameters given in Table~\ref{tab:input} and computed with SuperIso. \label{tab:obs}}
\label{Inputobservables}
\end{center}
\end{table}

\subsection{Results}

\begin{figure}[!t]
\begin{center}
\includegraphics[width=5.2cm]{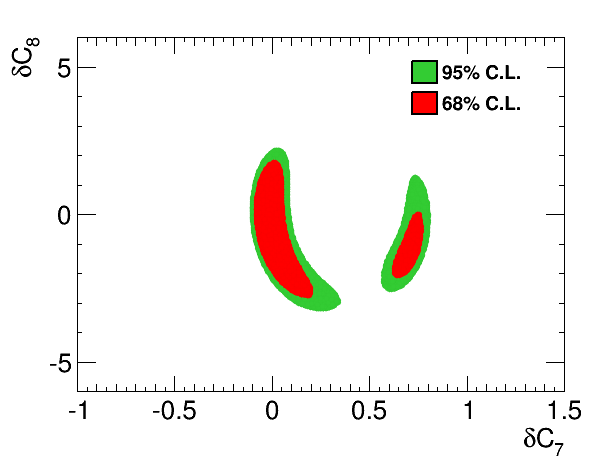}
\includegraphics[width=5.2cm]{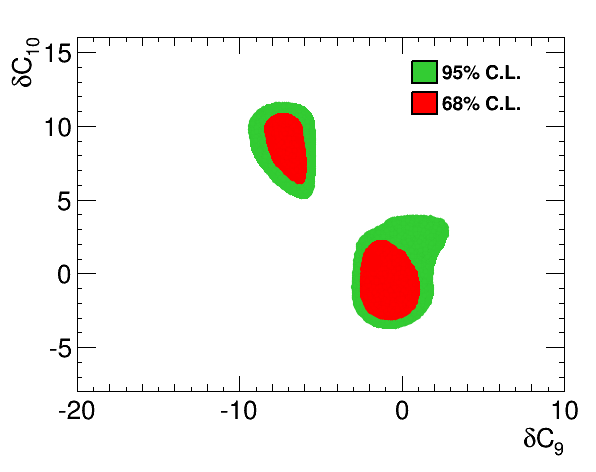}
\includegraphics[width=5.2cm]{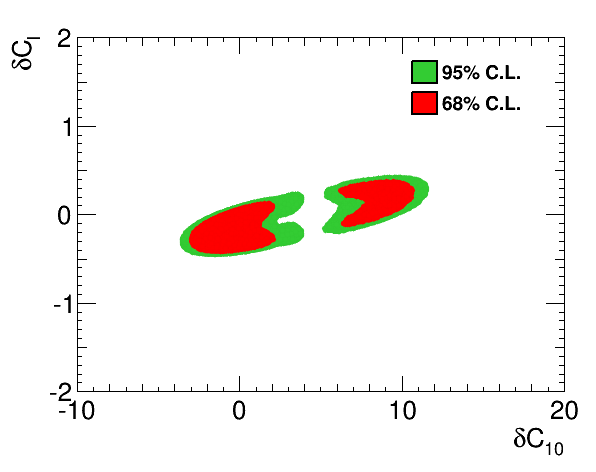}
\end{center}
\caption{Global fit to the NP contributions $\delta C_i$ in the MFV effective theory, at 68\% C.L. (red) and 95\% C.L. (green) using the 3 low-$q^2$ bins and the 2 high-$q^2$ bins of $B\to K^* \mu^+\mu^-$, and the other observables given in Table~\ref{tab:obs}.
\label{fig:MFV-all}}
\end{figure}

We first study the results of the global fit for the new physics contributions to the Wilson coefficients. For $B\to K^* \mu^+\mu^-$, we use the eight observables $P_1$, $P_2$, $P_4^{\prime}$, $P_5^{\prime}$, $P_6^{\prime}$, $P_8^{\prime}$, $F_L$ and the branching ratio in the three low $q^2$ bins and the two high $q^2$ bins. We also include BR($B \to X_s \gamma$), $\Delta_0(B \to K^* \gamma)$,
BR($B \to X_d \gamma$), BR($B_s \to \mu^+\mu^-$), BR($B_d \to \mu^+\mu^-$), BR($B \to X_s \mu^+\mu^-)_{q^2\in[1,6] \rm{GeV}^2}$ and BR($B \to X_s \mu^+\mu^-)_{q^2>14.4 \rm{GeV}^2}$, which in total amount to 47 observables in the fit, as given in Table~\ref{tab:obs}.
The 1 and 2$\sigma$ allowed regions are calculated as explained above, and the results for ($\delta C_7 , \delta C_8$), ($\delta C_9 , \delta C_{10}$) and ($\delta C_{10} , \delta C_l$) are presented in Fig.~\ref{fig:MFV-all}.

Compared to our previous analysis~\cite{Hurth:2012jn} where the new measurements for the optimised observables were not yet available, the allowed 68\% and 95\% regions are now smaller which shows the impact of the new measurements. $C_8$ is mostly constrained by $b\to s \gamma$ observables, $C_9$ and $C_{10}$ by $B\to K^* \mu^+\mu^-$, and $C_l$ by BR($B_s\to \mu^+\mu^-$). $C_7$ is constrained by most of the observables.
Similar to the previous results, two sets of solutions are possibles, corresponding to two separate zones in each plane, of which one contains the SM value of the Wilson coefficients (with $\delta C_i=0$) while the other corresponds to the opposite sign values. 

\begin{figure}[!h]
\begin{center}
\includegraphics[width=5.2cm]{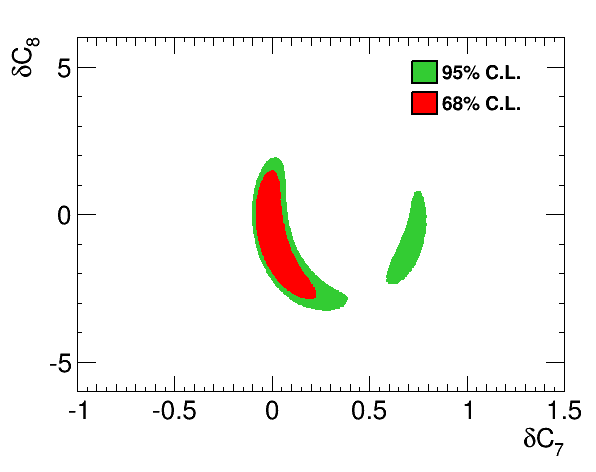}
\includegraphics[width=5.2cm]{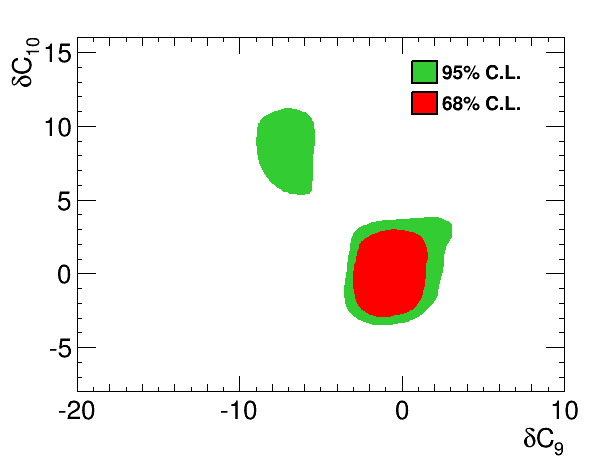}
\includegraphics[width=5.2cm]{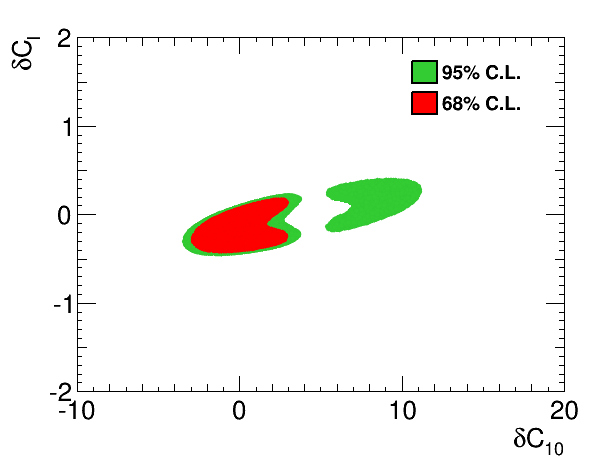}
\end{center}
\caption{Global fit to the NP coefficients $\delta C_i$ in the MFV effective theory, at 68\% C.L. (red) and 95\% C.L. (green) using the $q^2\in [1,6]$ GeV$^2$ bin and the 2 high-$q^2$ bins of $B\to K^* \mu^+\mu^-$, and the other observables given in Table~\ref{tab:obs}.
\label{fig:MFV-all16}}
\end{figure}

If instead of using the first three bins in the low-$q^2$ region for the $B\to K^* \mu^+\mu^-$ observables we use the $[1,6]$ bin, in which the deviations are smaller, the tension with the SM is reduced as can be seen in Fig.~\ref{fig:MFV-all16}. Comparing Figs.~\ref{fig:MFV-all} and \ref{fig:MFV-all16} reveals that using the $[1,6]$ bin, the zones including the SM still provide 1$\sigma$ acceptable solutions, while the other set could be in agreement with the experimental data only at the 2$\sigma$ level. 

\begin{figure}[!h]
\begin{center}
\includegraphics[width=5.2cm]{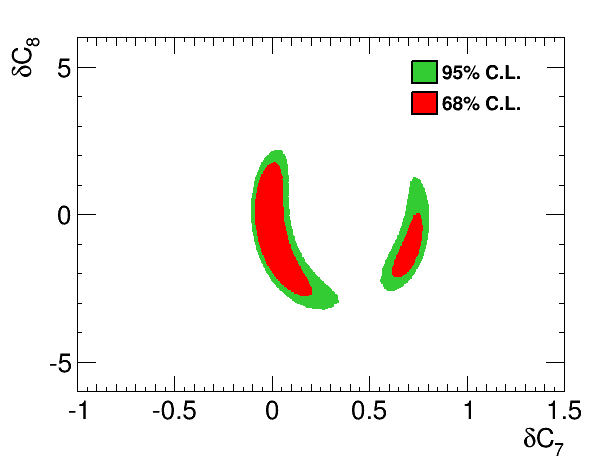}
\includegraphics[width=5.2cm]{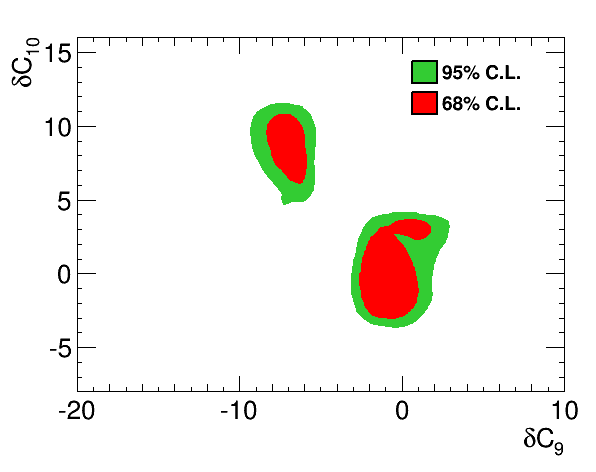}
\includegraphics[width=5.2cm]{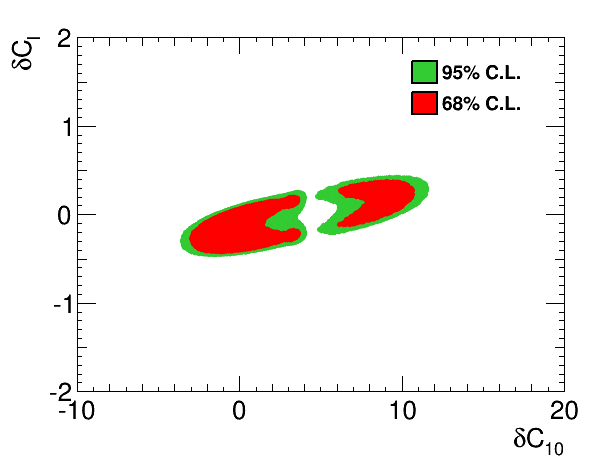}\\
\includegraphics[width=5.2cm]{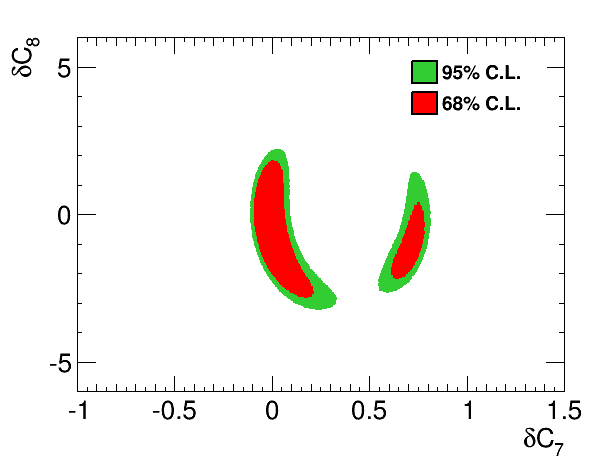}
\includegraphics[width=5.2cm]{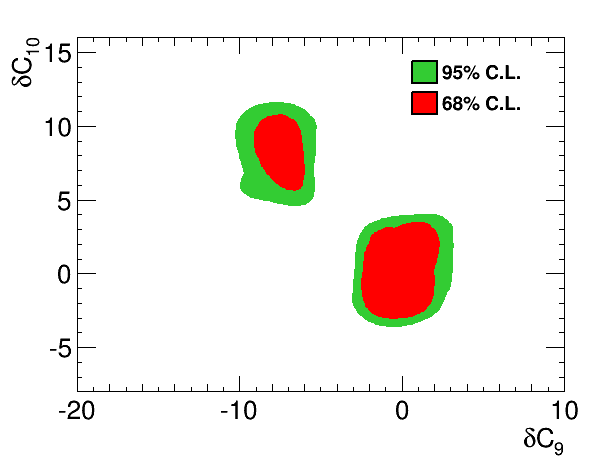}
\includegraphics[width=5.2cm]{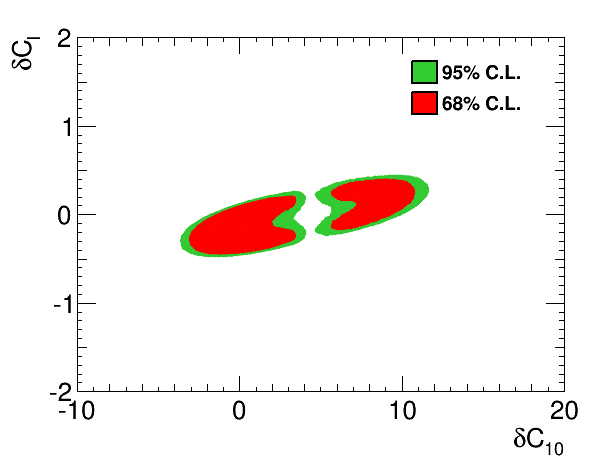}\\
\includegraphics[width=5.2cm]{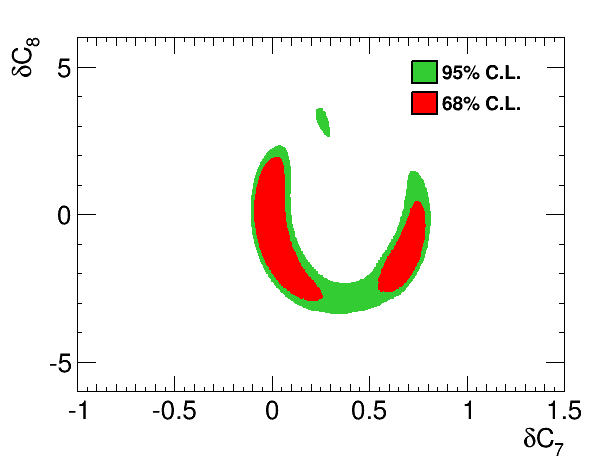}
\includegraphics[width=5.2cm]{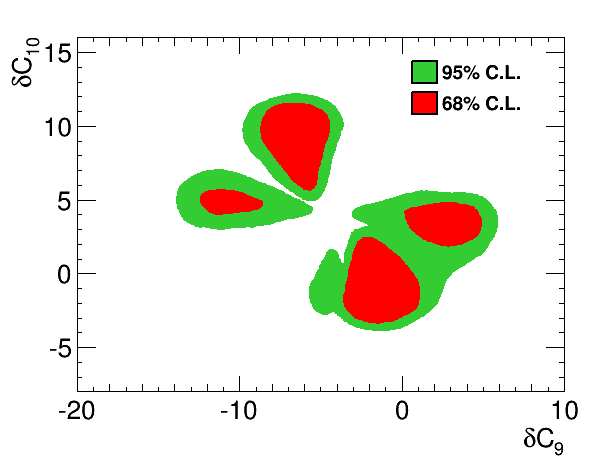}
\includegraphics[width=5.2cm]{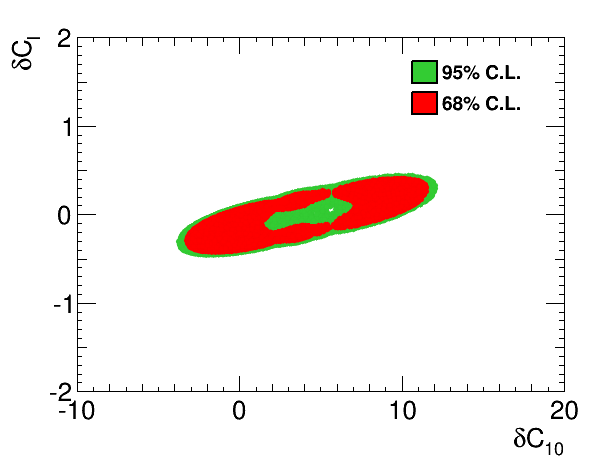}
\end{center}
\caption{Fit results, using all the observables except $P'_4$ (upper), except $P'_5$ (middle), and except $P_2$ (lower).
\label{fig:MFV-bounds}}
\end{figure}

To see the effect of the $B\to K^* \mu^+\mu^-$ observables which
present deviations with the SM predictions, namely $P_2$, $P'_4$ and
$P'_5$, we remove them one at the time from the global fit. The
difference with the results from the full fit is indicative of the
impact of the removed observable. The results are shown in
Fig.~\ref{fig:MFV-bounds}. As can be seen the impact of $P'_4$ and
$P'_5$ is rather mild, while removing $P_2$ makes a substantial change
in the 1 and 2$\sigma$ regions which are now enlarged. This shows the
important effect of $P_2$ on the global fit, which is mainly due to
the fact that the experimental measurement of $P_2$ is more
accurate.

\subsection{MFV predictions and bounds}

\begin{figure}[!t]
\begin{center}
\includegraphics[width=5.2cm]{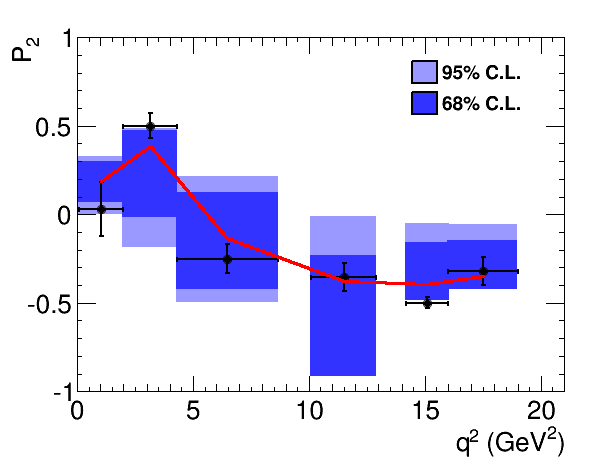}
\includegraphics[width=5.2cm]{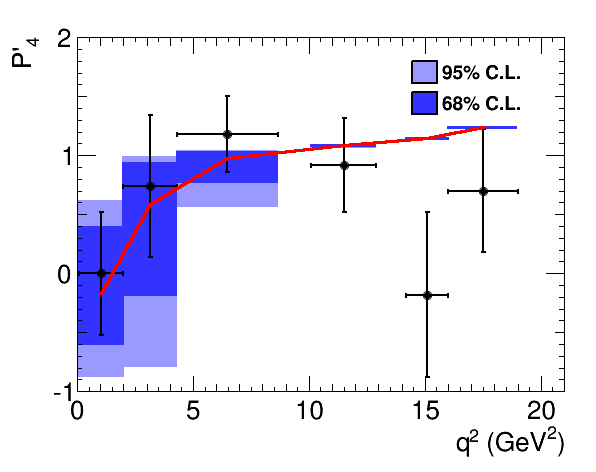}
\includegraphics[width=5.2cm]{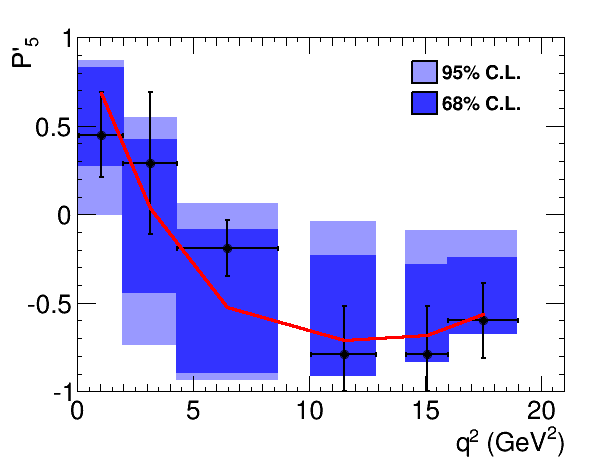}
\end{center}
\caption{MFV bounds for $P_2$ (left), $P'_4$ (centre) and $P'_5$ (right), using the results of the global fit at 68\% C.L. (dark blue) and 95\% C.L. (light blue), using the 3 low-$q^2$ bins and the 2 high-$q^2$ bins of $B\to K^* \mu^+\mu^-$ together with the other observables given in Table~\ref{tab:obs}. The red lines show the position of the best fit point.
\label{fig:MFV-binned}}
\end{figure}

The MFV solutions as a result of the global fit for $P_2$, $P'_4$ and $P'_5$ are displayed in Fig.~\ref{fig:MFV-binned} in each $q^2$ bin. The bands corresponding to the allowed 68 and 95\% C.L. regions are displayed in blue colours. The experimental results are also shown with black dots and error bars. It is remarkable that the 1$\sigma$ experimental errors overlap with the 1$\sigma$ range of the MFV predictions except for the $[14.18,16]$ bin in $P'_4$. It is however not possible to visualise the full $q^2$ distribution corresponding to each point in the fit. Therefore to guide the interpretation, we show the position of the best fit point throughout the bins with the red line, which shows a good overall agreement in all the bins, at least at the 2$\sigma$ level.  

\subsection*{Predictions for $P_2$, $P'_4$ and $P'_5$}

\begin{figure}[!t]
\begin{center}
\includegraphics[width=5.2cm]{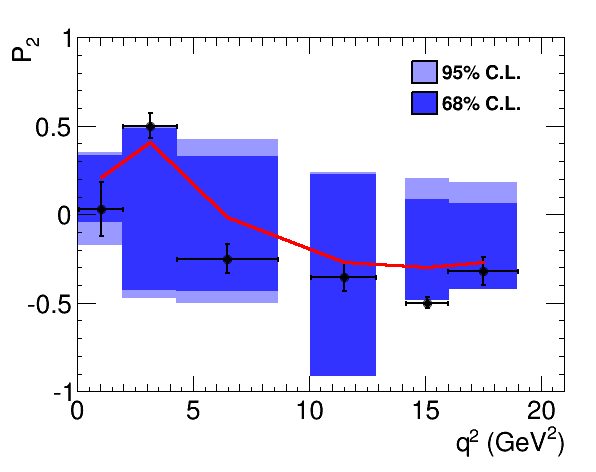}
\includegraphics[width=5.2cm]{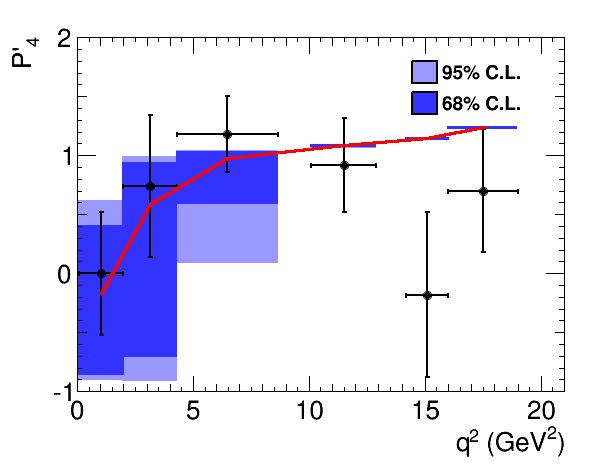}
\includegraphics[width=5.2cm]{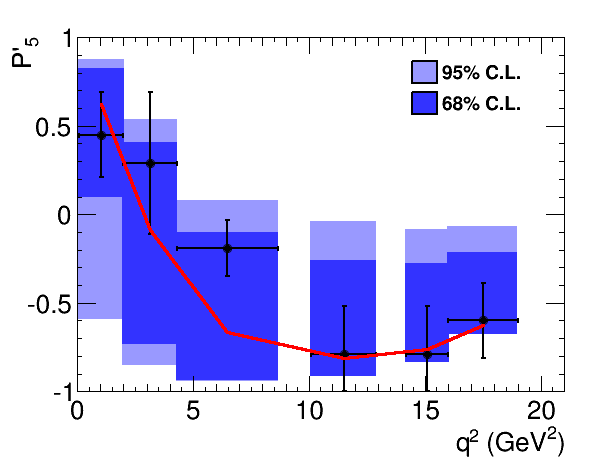}
\end{center}
\caption{MFV predictions for $P_2$ (left), $P'_4$ (centre) and $P'_5$ (right), obtained by removing $P_2$, $P'_4$ and $P'_5$ from the fit respectively, using the 3 low-$q^2$ bins and the 2 high-$q^2$ bins of $B\to K^* \mu^+\mu^-$ and the other observables given in Table~\ref{tab:obs}. The red lines show the position of the best fit point.
\label{fig:MFV-binned-Pi}}
\end{figure}

We can also check the MFV predictions for the observables which present deviations, namely $P_2$, $P'_4$ and $P'_5$. To make prediction for an observable, it is necessary to exclude that observable from the global fit. The predictions are shown in Fig.~\ref{fig:MFV-binned-Pi} for $P_2$, $P'_4$ and $P'_5$ respectively from left to right. Again the red line shows the predictions for the best fit point of the fit. The MFV predictions prove to be in good agreement with the experimental results, which are also shown in the figure.
 
The results presented in this section show that the overall agreement of the MFV solutions with the data is very good at the 2$\sigma$ level, and no new flavour structure is needed to explain the experimental results.

\section{Cross-check with the inclusive mode}

The inclusive mode $B\to X_s \ell^+ \ell^-$ 
provides complementary information to the exclusive $B\to K^* \mu^+\mu^-$ decay as already underlined in\cite{Hurth:2012jn}. 
This inclusive decay is theoretically well-explored. The NNLL QCD calculations for the 
branching ratio \cite{Bobeth:1999mk,Gambino:2003zm,Gorbahn:2004my,Asatryan:2001zw,Asatryan:2002iy,Ghinculov:2003qd,Ghinculov:2003bx} and the forward-backward asymmetry \cite{Ghinculov:2002pe,Asatrian:2002va,Asatrian:2003yk,Greub:2008cy,Bobeth:2003at,Huber:2008ak}
have been finalised some time ago by an effort of several research 
groups. Even electromagnetic corrections have been  already calculated~\cite{Bobeth:2003at,Huber:2005ig,Huber:2007vv}. The theoretical accuracy in the low-$q^2$ 
region is of the order of $10\%$~\cite{Huber:2007vv}.

We redo the global fit using only $B\to K^* \mu^+\mu^-$ observables and separately only $B\to X_s \ell^+\ell^-$ branching ratio and confront the results. Since the scalar contributions are neglected in the experimental results for the former, we also set them to zero in the following. 
For $B\to K^* \mu^+\mu^-$ we consider all the observables given in Table~\ref{tab:obs}. For $B\to X_s \ell^+\ell^-$, we combine the results from Belle and Babar for the branching ratio at low-$q^2$ and high-$q^2$.
In order to compare these two different sets of observables, we use now the $\Delta \chi^2$ fit method to obtain the exclusion plots of the Wilson coefficients. 
Indeed, the $\chi^2$ method we used in the previous section to test the overall consistency of the MFV hypothesis is not suitable for this comparison here because the exclusion plots would change if some less sensitive observables were removed from the fit. However, we have cross-checked and found very similar results using both methods.

\begin{figure}[t]
\begin{center}
\includegraphics[width=5.5cm]{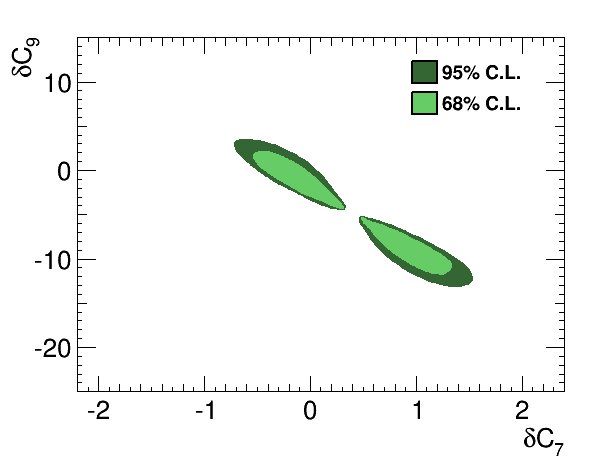}\quad\quad\includegraphics[width=5.5cm]{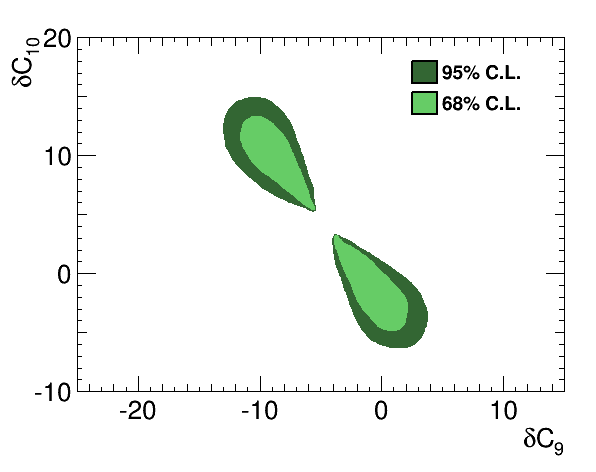}\\
\includegraphics[width=5.5cm]{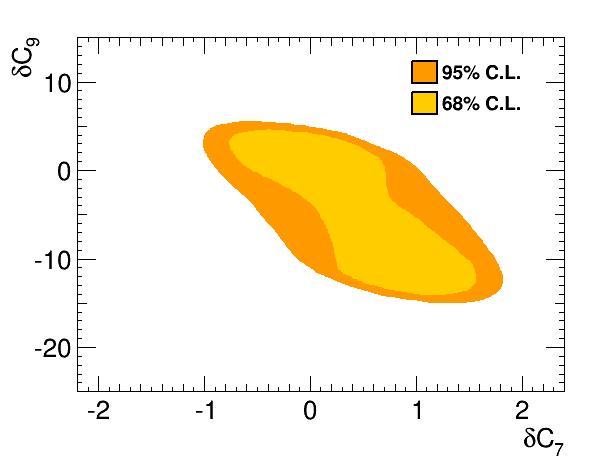}\quad\quad\includegraphics[width=5.5cm]{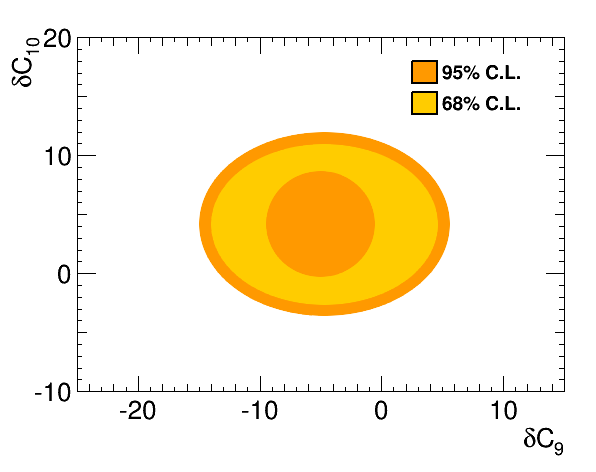}\\
\end{center}
\caption{$\Delta \chi^2$ fit results for the new physics contributions to $C_7$, $C_9$ and $C_{10}$, using only $B\to K^* \mu^+\mu^-$ observables in the low $q^2$ region (upper), using the current measurements of BR($B\to X_s \mu^+\mu^-$) at low and high $q^2$ (lower).\label{fig:bslla}}
\end{figure}

In Fig.~\ref{fig:bslla}, we illustrate the results of the $\Delta \chi^2$ fit for the relevant Wilson coefficients. The upper row shows the fit based on the exclusive 
($B \rightarrow K^* \ell^+\ell^-$) 
observables and the lower row the one based on the measurements of the inclusive ($B \to X_s\ell^+\ell^-$) branching ratio in the low- and high-$q^2$ regions. It is remarkable that the exclusion plot of the inclusive and the one of the exclusive modes are very similar and also compatible with each other. This is a nontrivial consistency check. 

However, unfortunately the latest measurements of the inclusive observables of the $B$ factories stem from 2004 in case of BaBar based on $89 \times 10^6 B \bar B$  
events~\cite{Aubert:2004it} from 2005 in case of Belle based on $152 \times 10^6 B\bar B$ events~\cite{Iwasaki:2005sy}. These numbers of events correspond to less than $30\%$ of the dataset available at the end of the $B$ factories. The analysis of the full datasets is expected to lead to a combined uncertainty of around $13\%$ for the measurements of the branching ratios~\cite{Kevin}. Thus, it will lead to even stronger constraints on the Wilson coefficients and to a more significant cross-check of the new physics hypothesis. 
 
\begin{figure}[t]
\begin{center}
\includegraphics[width=5.5cm]{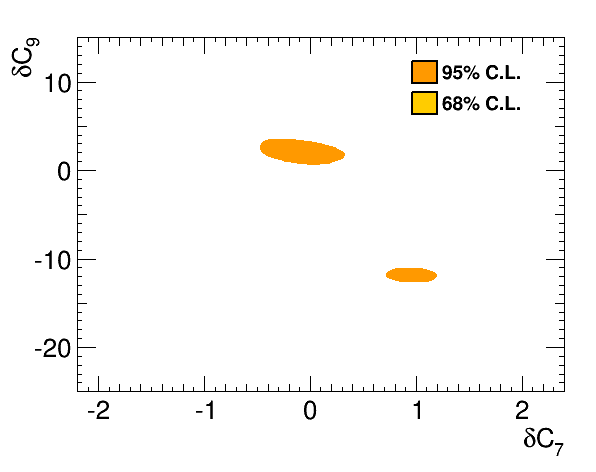}\quad\quad\includegraphics[width=5.5cm]{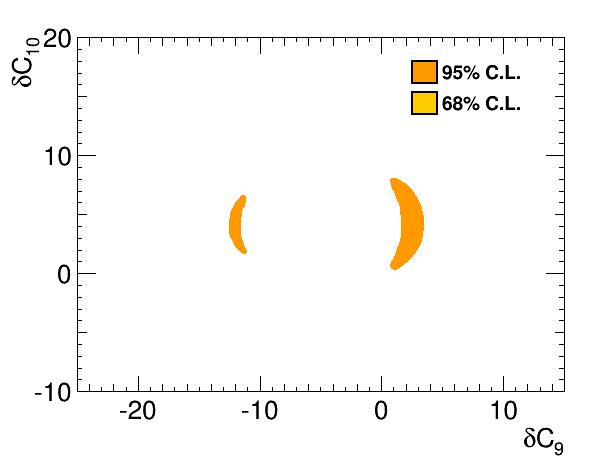}
\end{center}
\caption{$\chi^2$ fit results for the new physics contributions to $C_7$, $C_9$ and $C_{10}$, using the extrapolated measurements for BR($B\to X_s \mu^+\mu^-$) at low and high $q^2$ with the full Babar and Belle datasets assuming the central values as in the present measurements.\label{fig:bsllb}}
\end{figure}

\begin{figure}[t]
\begin{center}
\includegraphics[width=8.cm]{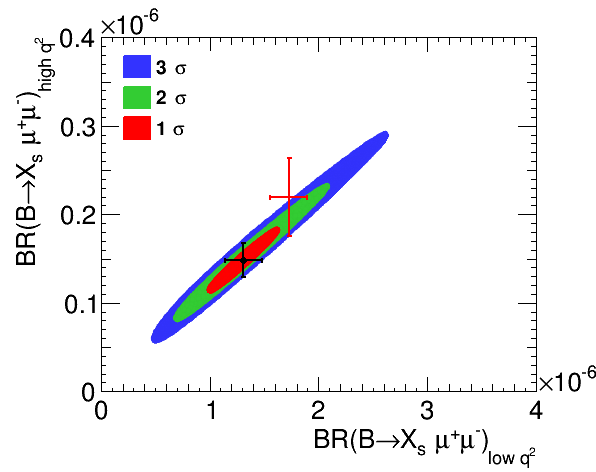}\end{center}
\caption{1, 2 and 3$\sigma$ ranges for the branching ratio at low- and high-$q^2$ within the model-independent analysis. 
Future measurement based on the full dataset of the $B $ factories ($13\%$ uncertainty) assuming the best-fit point of the model-independent analysis as central value (black) and the SM predictions (red/grey).\label{fig:bsllc}}
\end{figure}

Assuming the same central value as of the present measurements now with $13\%$ experimental errors for the final statistics of the $B$ factories, the $\chi^2$ fit results are 
very bad as Fig.~\ref{fig:bsllb} shows; one notices there is no compatibility at $68\%$ C.L. and the $95\%$ C.L. regions are very small. So a $\Delta \chi^2$-metrology does not make sense in this case. 
Therefore, let us illustrate the usefulness of a future measurement of the inclusive mode with the full dataset of the $B$ factories in another way: 
Based on the model-independent analysis of Ref.~\cite{Descotes-Genon:2013wba}, we predict the branching ratio at low- and high-$q^2$. In Fig.~\ref{fig:bsllc},
we show the 1, 2, and 3$\sigma$ ranges for these observables. In addition, we add the future measurements based on the full dataset with $13\%$ uncertainties assuming the best fit solution of the model-independent analysis of Ref.~\cite{Descotes-Genon:2013wba} as central value. These measurements are indicated by the black error bars. They should be compared with the theoretical SM predictions given by the red (grey) error bars. It is worth mentioning that the theory prediction for the high-$q^2$ region 
can be improved in the future by at least a factor 2.~\footnote{In fact, it is possible to drastically reduce the size of $1/m^2_b$ and $1/m^3_b$ power corrections to the integrated decay width, by normalising it to the semileptonic decay rate integrated over the {\it same} $q^2$ interval~\cite{Ligeti:2007sn}. This procedure will help reducing the uncertainties induced by the large power corrections to the decay width integrated over the high-$q^2$ region~\cite{Huber:2007vv}.}
Fig.~\ref{fig:bsllc} indicates that the future measurement of the inclusive branching ratios separates nicely from the SM prediction as the model-independent fit.

We can go one step further. In case the issue will be not resolved in the near future and more experimental accuracy is needed, there will be two dedicated flavour precision experiments:  
The upgrade of the LHCb experiment~\cite{Jacobsson:2013dla} will increase the integrated luminosity from 5 fb$^{-1}$ to 50 fb$^{-1}$, so the statistical uncertainties will get
decreased by a factor 3. However, the theory of exclusive modes will most probably not match this progress within the experimental measurements.
Moreover, there will be Super-$B$ factory Belle-II with a final integrated luminosity of 50 ab$^{-1}$. 
Fully inclusive measurements, i.e. those in which there are no a priori assumptions on the properties of the hadronic system accompanying the two final state leptons, can only be done at  such a high luminosity machine~\cite{belle2} by simultaneously reconstructing the two final state leptons and the accompanying recoiling $B$ meson produced in $\Upsilon(4S)$ decays. We follow here a recent analysis~\cite{Kevin2} of the expected total uncertainty on the partial decay width and the forward-backward asymmetry in several bins of dilepton mass-squared for the fully inclusive $B \to  X_s \ell^+\ell^-$ decays assuming a 50 ab$^{-1}$ total integrated luminosity. Based on some reasonable 
assumptions\footnote{The most important assumptions are the following: An overall efficiency of $2\%$ to reconstruct recoiling $B$ meson in either semileptonic or hadronic final states is assumed. After a tagged decay has been found, an efficiency of $60\%$  is assumed for the dilepton signal, which includes both geometric and reconstruction efficiencies. The dilepton mass-squared distribution assumed for signal decays is based on the theoretical prediction.  
Based on the experience at the first generation $B$ factories, signal-to-background ratios of order $\mathcal{O}(1)$ in the low-$q^2$ region can reasonably be expected for tagged $B$ events accompanied by two oppositely charged signal-side leptons. Significantly lower background rates can be expected in the high-$q^2$ region, and a signal-to-background ratio of $2.0$  is assumed there. Systematic uncertainties should be under good control using charmonium control samples decaying to the same final states as signal decays, and one  assigns a total systematic of $2\%$ in both the low- ($1<q^2<6$ GeV$^2$) and high-$q^2$ ($q^2 > 14.4$ GeV$^2$) regions.}
one finds a relative fractional uncertainty of $2.9\%$  ($4.1\%$) for the branching fraction in the low- (high-)$q^2$ region.  
\newpage
Moreover, a toy model including both signal and background contributions can be employed to estimate the $A_{FB}$-sensitivities\footnote{Within this toy model, signal distributions of the cosine of flavor-tagged lepton helicity angle at several values in the interval $[ -0.4 , 0.4]$ (for the normalised $A_{FB}$) are generated using the distribution expected for SM-like  $B \to K^* \ell^+\ell^-$ decays as a model. This is justified by noting that the distribution for fully inclusive SM signal events is not substantially different from that for SM 
$B \to K^* \ell^+ \ell^-$ decays. As before, backgrounds are assumed to occur at the same rate as signal in the lower $q^2$ region and half the signal rate in the upper range, with the additional assumption that there is no structure in the background helicity angle distribution. Central values and uncertainties for $A_{FB}$ are then derived from fitting background-subtracted toy distributions generated at several values of $A_{FB}$ to the function
Associated systematics are expected to be constrained using charmonium control samples, and one  assumes a $3\%$ value for the total angular analysis systematic uncertainty in any dilepton mass-squared interval.}. Within this toy model, one finds a total absolute uncertainty of 0.050 in the low-$q^2$ bin1 ($1<q^2<3.5$ GeV$^2$), 0.054 in the low-$q^2$ bin2  ($3.5<q^2<6$ GeV$^2$) and 0.058 in the high-$q^2$ interval ($q^2>14.4$ GeV$^2$)  for the {\it normalised}  $A_{FB}$.
\begin{figure}[t]
\begin{center}
\includegraphics[width=8.cm]{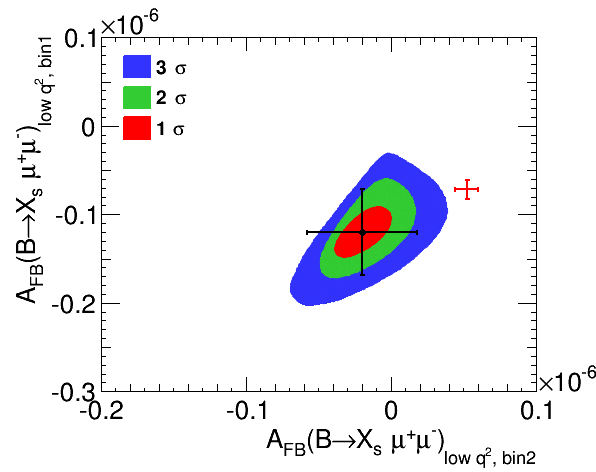}
\end{center}
\caption{1, 2 and 3$\sigma$ ranges for the {\it unnormalised} forward-backward asymmetry in bin 1 ($1<q^2<3.5$ GeV$^2$) and in bin 2 ($3.5<q^2<6$ GeV$^2$) within the model-independent analysis. Future measurement at the high-luminosity Belle-II Super-$B$-Factory assuming the best-fit point of the 
model-independent analysis as central value (black) and the SM predictions (red/grey).\label{fig:bslld}}
\end{figure}

With this expected performance of the Belle-II experiment, the measurement of the branching ratios will be possible with much smaller uncertainties. In Fig.~\ref{fig:bsllc} the experimental
error bars will get smaller by more than a factor 2 with Belle-II. 

Also the future measurement of the forward-backward asymmetry at Belle-II will  allow to separate the potential new physics measurement from the SM prediction in a significant way as shown in Fig~\ref{fig:bslld}. Note the zero of the forward backward asymmetry is pushed to higher values with the best fit solution of the model-independent analysis:
the 1$\sigma$-interval from the model-independent fit at NLO is: $4.74-5.51$.  This implies that the integrated forward-backward asymmetry of the second low-$q^2$ bin 
is also negative.

\newpage 

\section{Conclusion}
The LHCb collaboration has presented an angular analysis of the decay mode $B \to K^* \mu^+\mu^-$ based on 1 fb$^{-1}$.
LHCb has found a 4.0$\sigma$ local discrepancy in one of the low-$q^2$ bins for one of the angular observables. This deviation together with other smaller deviations in the angular analysis can be consistently described by a smaller $C_9$ Wilson coefficient, together with a less significant contribution of a non-zero $C_9^{\prime}$. 

Clearly, this exciting LHCb result calls for a better understanding of the power correction to the decay mode. They lead to the largest part of the theoretical uncertainty because they are undetermined within the QCD-factorisation approach which is the state of the art method for the 
low-$q^2$ region. However, there are soft arguments that such uncertainties are of the order of $10\%$. 

Possible cross-checks with other observables are interesting. First, there are the corresponding angular observables at the high-$q^2$ region which are based on different theoretical methods with lower uncertainties. Second, we made manifest that these final measurements of Babar and Belle allow for important cross-checks of the new physics hypothesis.  

We also showed that SuperLHCb and Belle-II might play a role if the new physics signals need more experimental accuracy.

Finally, assuming that the LHCb anomaly is a hint of NP, we showed within a detailed MFV analysis that no new flavour structures are needed.

\newpage
\appendix

\section*{Appendix: correlation matrices}

We provide below the correlation matrices for each bin~\cite{Serra}. They have been estimated by using a toy Monte Carlo technique. 

\begin{table}[!htb]
\begin{center}
\begin{tabular}{|c|cccccccc|}
\hline
& $P_1$ & $P_2$ & $P_4^{\prime}$ & $P_5^{\prime}$ & $P_6^{\prime}$  & $P_8^{\prime}$ &$F_L$ & $A_{FB}$\\
\hline
$P_1$ &1.00 & -0.01 & 0.05 & 0.09 & -0.00 & 0.04 & -0.02 & 0.01 \\ 
$P_2$ &-0.01 & 1.00 & 0.08 & 0.02 & 0.02 & -0.02 & 0.07 & -0.97 \\ 
$P_4^{\prime}$ & 0.05 & 0.08 & 1.00 & 0.05 & -0.06 & -0.04 & 0.00 & -0.08 \\ 
$P_5^{\prime}$ & 0.09 & 0.02 & 0.05 & 1.00 & -0.08 & 0.02 & -0.08 & -0.01 \\ 
$P_6^{\prime}$  & -0.00 & 0.02 & -0.06 & -0.08 & 1.00 & -0.01 & -0.06 & -0.02 \\ 
$P_8^{\prime}$ & 0.04 & -0.02 & -0.04 & 0.02 & -0.01 & 1.00 & -0.05 & 0.02 \\ 
$F_L$ & -0.02 & 0.07 & 0.00 & -0.08 & -0.06 & -0.05 & 1.00 & -0.07 \\ 
$A_{FB}$ & 0.01 & -0.97 & -0.08 & -0.01 & -0.02 & 0.02 & -0.07 & 1.00 \\
\hline
\end{tabular}
\caption{Correlation matrix for the $q^2$ region $[0.1,2.0]$ GeV$^2$ estimated by using a toy Monte Carlo technique.\label{tab:correlations_bin1}}
\end{center}
\end{table}

\begin{table}[!htb]
\begin{center}
\begin{tabular}{|c|cccccccc|}
\hline
& $P_1$ & $P_2$ & $P_4^{\prime}$ & $P_5^{\prime}$ & $P_6^{\prime}$  & $P_8^{\prime}$ &$F_L$ & $A_{FB}$\\
\hline
$P_1$ &1.00 & 0.02 & 0.08 & 0.05 & -0.03 & -0.00 & -0.03 & -0.02 \\ 
$P_2$ &0.02 & 1.00 & 0.02 & 0.11 & -0.01 & -0.01 & 0.10 & -0.73 \\ 
$P_4^{\prime}$ & 0.08 & 0.02 & 1.00 & 0.28 & 0.02 & -0.00 & 0.03 & -0.02 \\ 
$P_5^{\prime}$ & 0.05 & 0.11 & 0.28 & 1.00 & 0.02 & -0.05 & 0.03 & -0.11 \\ 
$P_6^{\prime}$  & -0.03 & -0.01 & 0.02 & 0.02 & 1.00 & 0.22 & -0.03 & 0.01 \\ 
$P_8^{\prime}$ & -0.00 & -0.01 & -0.00 & -0.05 & 0.22 & 1.00 & -0.03 & 0.01 \\ 
$F_L$ & -0.03 & 0.10 & 0.03 & 0.03 & -0.03 & -0.03 & 1.00 & -0.10 \\ 
$A_{FB}$ & -0.02 & -0.73 & -0.02 & -0.11 & 0.01 & 0.01 & -0.10 & 1.00 \\
\hline
\end{tabular}
\caption{Correlation matrix for the $q^2$ region $[2.0,4.3]$ GeV$^2$ estimated by using a toy Monte Carlo technique.\label{tab:correlations_bin2}}
\end{center}
\end{table}

\begin{table}[!htb]
\begin{center}
\begin{tabular}{|c|cccccccc|}
\hline
& $P_1$ & $P_2$ & $P_4^{\prime}$ & $P_5^{\prime}$ & $P_6^{\prime}$  & $P_8^{\prime}$ &$F_L$ & $A_{FB}$\\
\hline
$P_1$ &1.00 & -0.02 & 0.14 & -0.03 & -0.03 & 0.04 & 0.07 & 0.02 \\ 
$P_2$ &-0.02 & 1.00 & 0.03 & 0.18 & -0.07 & -0.02 & -0.13 & -0.97 \\ 
$P_4^{\prime}$ & 0.14 & 0.03 & 1.00 & -0.16 & -0.05 & 0.03 & -0.04 & -0.03 \\ 
$P_5^{\prime}$ & -0.03 & 0.18 & -0.16 & 1.00 & 0.04 & 0.01 & 0.02 & -0.18 \\ 
$P_6^{\prime}$  & -0.03 & -0.07 & -0.05 & 0.04 & 1.00 & -0.14 & -0.01 & 0.07 \\ 
$P_8^{\prime}$ & 0.04 & -0.02 & 0.03 & 0.01 & -0.14 & 1.00 & 0.01 & 0.02 \\ 
$F_L$ & 0.07 & -0.13 & -0.04 & 0.02 & -0.01 & 0.01 & 1.00 & 0.13 \\ 
$A_{FB}$ & 0.02 & -0.97 & -0.03 & -0.18 & 0.07 & 0.02 & 0.13 & 1.00 \\ 
\hline
\end{tabular}
\caption{Correlation matrix for the $q^2$ region $[4.3,8.68]$ GeV$^2$ estimated by using a toy Monte Carlo technique.}
\end{center}
\end{table}

\begin{table}[!htb]
\begin{center}
\begin{tabular}{|c|cccccccc|}
\hline
& $P_1$ & $P_2$ & $P_4^{\prime}$ & $P_5^{\prime}$ & $P_6^{\prime}$  & $P_8^{\prime}$ &$F_L$ & $A_{FB}$\\
\hline
$P_1$ &1.00 & -0.02 & 0.03 & -0.22 & -0.08 & -0.03 & -0.14 & 0.01 \\ 
$P_2$ &-0.02 & 1.00 & -0.14 & 0.13 & -0.04 & 0.00 & -0.13 & -0.93 \\ 
$P_4^{\prime}$ & 0.03 & -0.14 & 1.00 & -0.14 & 0.02 & 0.04 & -0.07 & 0.13 \\ 
$P_5^{\prime}$ & -0.22 & 0.13 & -0.14 & 1.00 & -0.06 & 0.03 & -0.07 & -0.12 \\ 
$P_6^{\prime}$  & -0.08 & -0.04 & 0.02 & -0.06 & 1.00 & -0.20 & -0.03 & 0.04 \\ 
$P_8^{\prime}$ & -0.03 & 0.00 & 0.04 & 0.03 & -0.20 & 1.00 & 0.03 & -0.01 \\ 
$F_L$ & -0.14 & -0.13 & -0.07 & -0.07 & -0.03 & 0.03 & 1.00 & 0.13 \\ 
$A_{FB}$ & 0.01 & -0.93 & 0.13 & -0.12 & 0.04 & -0.01 & 0.13 & 1.00 \\
\hline
\end{tabular}
\caption{Correlation matrix for the $q^2$ region $[10.09,12.90]$ GeV$^2$ estimated by using a toy Monte Carlo technique.\label{tab:correlations_bin4}}
\end{center}
\end{table}

\begin{table}[!htb]
\begin{center}
\begin{tabular}{|c|cccccccc|}
\hline
& $P_1$ & $P_2$ & $P_4^{\prime}$ & $P_5^{\prime}$ & $P_6^{\prime}$  & $P_8^{\prime}$ &$F_L$ & $A_{FB}$\\
\hline
$P_1$ &1.00 & 0.00 & 0.00 & -0.21 & 0.02 & -0.05 & 0.08 & -0.00 \\ 
$P_2$ &0.00 & 1.00 & -0.03 & -0.04 & 0.01 & -0.00 & -0.02 & -0.95 \\ 
$P_4^{\prime}$ & 0.00 & -0.03 & 1.00 & -0.36 & 0.03 & -0.04 & 0.00 & 0.03 \\ 
$P_5^{\prime}$ & -0.21 & -0.04 & -0.36 & 1.00 & -0.02 & 0.02 & 0.02 & 0.04 \\ 
$P_6^{\prime}$  & 0.02 & 0.01 & 0.03 & -0.02 & 1.00 & -0.40 & 0.05 & -0.01 \\ 
$P_8^{\prime}$ & -0.05 & -0.00 & -0.04 & 0.02 & -0.40 & 1.00 & 0.00 & 0.00 \\ 
$F_L$ & 0.08 & -0.02 & 0.00 & 0.02 & 0.05 & 0.00 & 1.00 & 0.02 \\ 
$A_{FB}$ & -0.00 & -0.95 & 0.03 & 0.04 & -0.01 & 0.00 & 0.02 & 1.00 \\ 
\hline
\end{tabular}
\caption{Correlation matrix for the $q^2$ region $[14.18,16.00]$ GeV$^2$ estimated by using a toy Monte Carlo technique.\label{tab:correlations_bin5}}
\end{center}
\end{table}

\begin{table}[!htb]
\begin{center}
\begin{tabular}{|c|cccccccc|}
\hline
& $P_1$ & $P_2$ & $P_4^{\prime}$ & $P_5^{\prime}$ & $P_6^{\prime}$  & $P_8^{\prime}$ &$F_L$ & $A_{FB}$\\
\hline
$P_1$ &1.00 & 0.01 & -0.05 & -0.22 & -0.03 & 0.04 & -0.05 & -0.01 \\ 
$P_2$ &0.01 & 1.00 & -0.17 & -0.12 & 0.05 & 0.03 & -0.14 & -0.95 \\ 
$P_4^{\prime}$ & -0.05 & -0.17 & 1.00 & -0.39 & 0.02 & -0.05 & -0.05 & 0.16 \\ 
$P_5^{\prime}$ & -0.22 & -0.12 & -0.39 & 1.00 & 0.04 & 0.00 & -0.04 & 0.12 \\ 
$P_6^{\prime}$  & -0.03 & 0.05 & 0.02 & 0.04 & 1.00 & -0.37 & -0.03 & -0.05 \\ 
$P_8^{\prime}$ & 0.04 & 0.03 & -0.05 & 0.00 & -0.37 & 1.00 & -0.02 & -0.03 \\ 
$F_L$ & -0.05 & -0.14 & -0.05 & -0.04 & -0.03 & -0.02 & 1.00 & 0.14 \\ 
$A_{FB}$ & -0.01 & -0.95 & 0.16 & 0.12 & -0.05 & -0.03 & 0.14 & 1.00 \\
\hline
\end{tabular}
\caption{Correlation matrix for the $q^2$ region $[16.0,19.0]$ GeV$^2$ estimated by using a toy Monte Carlo technique.\label{tab:correlations_bin6}}
\end{center}
\end{table}

\begin{table}[!htb]
\begin{center}
\begin{tabular}{|c|cccccccc|}
\hline
& $P_1$ & $P_2$ & $P_4^{\prime}$ & $P_5^{\prime}$ & $P_6^{\prime}$ & $P_8^{\prime}$ & $F_L$ & $A_{FB}$\\
\hline
$P_1$ &1.00 & 0.09 & 0.10 & -0.16 & -0.01 & -0.01 & -0.09 & -0.08 \\ 
$P_2$ &0.09 & 1.00 & -0.27 & 0.11 & 0.01 & -0.03 & 0.09 & -0.77 \\ 
$P_4^{\prime}$ & 0.10 & -0.27 & 1.00 & 0.28 & -0.01 & 0.01 & 0.01 & 0.28 \\ 
$P_5^{\prime}$ & -0.16 & 0.11 & 0.28 & 1.00 & -0.03 & -0.01 & -0.11 & -0.10 \\ 
$P_6^{\prime}$  & -0.01 & 0.01 & -0.01 & -0.03 & 1.00 & 0.26 & -0.01 & -0.01 \\ 
$P_8^{\prime}$ & -0.01 & -0.03 & 0.01 & -0.01 & 0.26 & 1.00 & -0.05 & 0.03 \\ 
$F_L$ & -0.09 & 0.09 & 0.01 & -0.11 & -0.01 & -0.05 & 1.00 & -0.09 \\ 
$A_{FB}$ & -0.08 & -0.77 & 0.28 & -0.10 & -0.01 & 0.03 & -0.09 & 1.00 \\
\hline
\end{tabular}
\caption{Correlation matrix for the $q^2$ region $[1.0,6.0]$ GeV$^2$ estimated by using a toy Monte Carlo technique.\label{tab:correlations_bin7}}
\end{center}
\end{table}

\section*{Acknowledgement} 

TH thanks the CERN theory group for its hospitality during his regular visits to CERN where part of this work was written. The authors are very grateful to Nicola Serra for providing us with the experimental correlation matrices for $B \to K^* \mu^+ \mu^-$ observables and for his input. TH is very grateful to Jerome Charles for innumerable comments and discussions, and FM to Marco Battaglia for useful discussions. Finally, the authors are very grateful to Kevin Flood for his analysis of the Belle-II performance. 


\end{document}